\documentclass[11pt]{article}
\usepackage{amssymb,amsmath,amsthm}	% Mathematische Symbole
\usepackage{a4wide}
\usepackage[utf8x]{luainputenc}               	% Eingabe von Umlauten und ß
\usepackage{graphicx}      										% Grafiken einbinden
\usepackage{cite}
\usepackage[numbers]{natbib}
\usepackage{dsfont}
\usepackage{hyperref}
\usepackage{multirow}
\usepackage{subcaption}
\usepackage{amsmath}
\usepackage{xspace}
\usepackage{cleveref}
\usepackage{xcolor}
\usepackage{ulem}

\newcommand{\klabsatz}{\\[8pt]}          % kleiner Absatz
\newtheorem{definition}{Definition}[]

\newtheorem{theorem}{Theorem}[]
\newtheorem{remark}{Remark}[]

\newtheorem{proposition}{Proposition}[]

\newcommand{\R}{\ensuremath{\mathbb{R}}}

\newcommand{\N}{\ensuremath{\mathbb{N}}}

\newcommand{\p}[2]{\ensuremath{\frac{\partial #1}{\partial #2 }}}
\newcommand{\bs}{\ensuremath{\boldsymbol}}
\newcommand{\diag}{\mathop{\mathrm{diag}}}
\newcommand{\dist}{\mathop{\mathrm{dist}}}
\usepackage{lineno}

\begin{document}

%\linenumbers
	%-- TITEL ----------------------------------------------------------%

	\title{\textcolor{black}{Robust Mathematical Formulation and Probabilistic Description of Agent-Based Computational Economic Market Models}}

	\author{ Maximilian Beikirch\footnote{RWTH Aachen University, Templergraben 55, 52056 Aachen, Germany}~\footnote{ORCiD IDs: Maximilian Beikirch: 0000-0001-6055-4089, Simon Cramer: 0000-0002-6342-8157, Philipp Otte: 0000-0002-1586-2274, Emma Pabich: 0000-0002-0514-7402, Torsten Trimborn:  0000-0001-5134-7643}, Simon Cramer\footnote{Chair of Production Metrology and Quality Management at WZL, RWTH Aachen University}~\footnotemark[2],
	 Martin Frank\footnote{Karlsruhe Institute of Technology, Steinbuch Center for Computing, Hermann-von-Helmholtz-Platz 1, 76344 Eggenstein-Leopoldshafen, Germany},\\
	Philipp Otte\footnote{Forschungszentrum J\"ulich GmbH, Institute for Advanced Simulation, J\"ulich Supercomputing Centre, 52425 J\"ulich, Germany} \footnotemark[2], Emma Pabich\footnote{Institute for Data Science in Mechanical Engineering, RWTH Aachen University}~\footnotemark[2], Torsten Trimborn\footnote{NRW.BANK, Kavalleriestra\ss e  22, 40213 D\"usseldorf, Germany}~\footnotemark[2]~\footnote{Corresponding author: torsten.trimborn@nrwbank.de}}

\maketitle

   %-- INHALTSVERZEICHNIS ----------------------------------------------------------%

%	\tableofcontents

	%-- EINLEITUNG ----------------------------------------------------------%

\begin{abstract}
\textcolor{black}{In science and especially in economics, agent-based modeling has become a widely used modeling approach.
These models are often formulated as a large system of difference equations.
In this study, we discuss two aspects, numerical modeling and the probabilistic description for two agent-based computational economic market models: the Levy-Levy-Solomon model and the Franke-Westerhoff model.
We derive time-continuous formulations of both models, and in particular we discuss the impact of the time-scaling on the model behavior for the Levy-Levy-Solomon model.
For the Franke-Westerhoff model, we proof that a constraint required in the original model is not necessary for stability of the time-continuous model.
It is shown that a semi-implicit discretization of the time-continuous system preserves this unconditional stability. 
In addition, this semi-implicit discretization can be computed at cost comparable to the original model.
Furthermore, we discuss possible probabilistic descriptions of time continuous agent-based computational economic market models.
Especially, we present the potential advantages of kinetic theory in order to derive mesoscopic desciptions of agent-based models.
Exemplified, we show two probabilistic descriptions of the Levy-Levy-Solomon and Franke-Westerhoff model. }
 \end{abstract}

\section{Introduction}
Agent-based computational economic market (ABCEM) models have become a popular modeling tool in economic research. They can be regarded as large dynamical systems formulated as difference equations. The model by Stigler \cite{stigler1964theory} may be seen as the first  ABCEM model but generally the model by Kim and Markowitz \cite{kim1989investment, samanidou2007agent} is referred to as the first modern ABCEM model. ABCEM models are a notable class in the research field econophysics. The general  goal of ABCEM models is to reproduce persistent statistical features present in financial data all over the world which are known as stylized facts. Possible research questions include: evaluate the kind of stylized facts microscopic agent dynamics create; and estimate the impact of regulations on a financial market. Thanks to Monte Carlo simulations, it is possible to study the time evolution of statistical quantities such as the wealth distribution or the stock return distribution.  The importance of agent-based modeling in economics has been emphasized by several authors \cite{farmer2009economy, hommes2006heterogeneous, tesfatsion2002agent}. \\ \\
The drawback of this procedure is that these empirical results solely rely on computational experiments. In addition, the sensitivity of many ABCEM models w.r.t. their parameters, observed in a tendency towards blow ups for particular choices of model parameters, motivated this work.
The deeper study of these phenomena has revealed that this behavior often originates from the time stepping scheme of difference equations which can be viewed as explicit Euler discretizations with a normalized time step of one. It is  a well known property of the explicit Euler method to perform poorly in the case of stiff differential equations. \textcolor{black}{Here, we consider an ODE/SDE stiff if applying an explicit Euler-type discretization (explicit Euler for ODEs and Euler-Maruyama for SDEs) enforces a very small timestep for the timestepping to be stable.} For this reason, we believe that a time continuous formulation of difference equations helps understanding several numerical issues. Albeit, we have to recognize that starting from a difference equation, the continuous formulation, respectively the continuum limit, is not uniquely defined. To our knowledge, this approach is the first work within the ABCEM community addressing this important issue. \textcolor{black}{ In economic literature in general, the connection between time-continuous and time-discrete models has been studied in the context of volatility models in finance by Nelson \cite{nelson1990arch} and for macroeconomic models by Turnovsky  \cite{turnovsky1977formulation}} \\ \\
For these reasons, the goal of this paper is to study the continuous formulation of ABCEM models and to present strategies for a robust mathematical formulation of ABCEM models.
\textcolor{black}{ Furthermore, we emphasize that a  time continuous dynamical system may be translated to a mesoscopic description modeled using partial differential equations (PDEs) \cite{lapeyre2003introduction, pareschi2013interacting}. By probabilistic description we refer to a partial differential equation (PDE) whose solution is the density function of a probability law. 
The limit process leading from microscopic dynamics to a mesoscopic description is at the heart of kinetic theory which has been successfully applied to several ABCEM models in the past \cite{albi2014kinetic, albi2017recent, trimborn2018mean, trimborn2017portfolio, trimborn2019macroscopic, maldarella2012kinetic}.  In comparison to large dynamical systems the dimension of the PDE solution is usually low dimensional and can be conveniently analyzed by mathematical tools.} Clearly these previously discussed aspects apply to a wider class of agent-based models. In this work we analyze the following issues:
\begin{enumerate}
\item  \textbf{Continuous Formulation and Limit}: We discuss the connections of difference equations and differential equations. In particular, we emphasize that the continuum formulation of a difference equation is not unique. In addition, we claim that a continuous models, such as stochastic and ordinary differential equations, have several advantages in comparison to difference equations. First of all numerical methods can be conveniently applied. In addition a continuous formulation enables to pass to a probabilistic description.
\item \textbf{Numerical Discretization and Solver}: Numerical discretization include time discretization schemes of stochastic and ordinary equations as well as discretizations of differential occuring within formulas. Numerical solvers include numerical solvers for non-linear equations. We highlight that the choice of numerical discretizations and solvers is paramount, e.g. for the solution of stiff ordinary differential equations implicit methods are beneficial.
\item\textcolor{black}{ \textbf{Probabilistic Description:} Starting from a time continuous ABCEM model we discuss several techniques to derive mesoscopic PDE models, starting with microscopic agent dynamics. We present the classical Feynman-Kac formula and give an introduction to kinetic theory. Especially, we want to point out that a mesoscopic model enables us to study the long time behavior or even to compute the solution. }
\end{enumerate}
The choice to formulate an ABCEM model continuously or as a difference equation is a questions of modeling.
 Difference equations are able to model a time evolution of quantities with a fixed time lag. In contrast, a time continuous model, e.g.
an ordinary differential equation (ODE) or stochastic differential equation (SDE), may be discretized using various time steps and is thus suitable to model the same quantity resolving different time scales. 
Any difference equations can be interpreted as a scaled Euler-type discretization of a time continuous differential equation. 
Note that starting from a difference equation, the time continuous formulation is not unique. Furthermore, the advantage of a time continuous formulation is that it may be used to explain instabilities on the level of the difference equations and thus to guide the choice of appropriate discretization schemes. As pointed out before a time continuous dynamical system enables to derive PDE models.
\textcolor{black}{These mesoscopic models might be solely a different description of the original microscopic model or an approximation of the original system. Nevertheless a PDE model has the advantage that it is accessible to rigorous analysis and to prove for example interconnections between agents' modeling and the appearance of stylized facts.} \\ \\
 In this study we exemplarily discuss the continuous formulation, continuous limit and mesoscopic description of the Levy-Levy-Solomon model \cite{levy1994microscopic} and Franke-Westerhoff model \cite{franke2009validation}. 
The Levy-Levy-Solomon model is one of the most influential ABCEM models and an early example of an ABCEM model in general \cite{samanidou2007agent}. The Levy-Levy-Solomon model considers the wealth evolution of agents and the stock price evolution. Furthermore, each agent has to decide in each time step on the optimal asset allocation between the stocks and the asset class bonds. The stock price is fixed in each time step by the clearance mechanism that perfectly matches supply and demand and can be consequently seen as a rational market. The authors claimed that their model is able to reproduce several stylized facts from financial markets such as fat-tails in asset returns. It has been documented by Zschischang and Lux \cite{zschischang2001some} that the stock price returns are normally distributed and that the model exhibits finite-size effects.
 In comparison to the Levy-Levy-Solomon model, the Franke-Westerhoff model is rather recent and has been first introduced in 2009 \cite{franke2009validation}. The model tracks the time evolution of two agent groups and not of an individual agent as in the Levy-Levy-Solomon model. The stock price is modeled as a stochastic difference equation and is thus modeled as a disequilibrium model. The Franke-Westerhoff model is fully described by a system of three difference equations. It has been documented that the Franke-Westerhoff model is able to reproduce several stylized facts.
The reason to choose the Franke-Westerhoff model in this study is on the one hand the simplicity of this model and on the other hand the prototypical nature of this model especially w.r.t. the stock price update mechanism. The Levy-Levy-Solomon model was selected not only because of the popularity but as well in virtue of the wealth evolution of agents. 
The key point in ABCEM models is the interaction among agents. The Levy-Levy-Solomon considers direct interactions of agents, which are hidden in the definition of the stock price. The Franke-Westerhoff model considers two \textit{representative} agents which interact implicitly through the stock price equation and the switching probabilities. \\ \\ 
The paper is structured as follows.  In the next section, we discuss the connection of difference equations and time continuous differential equations in the context of disequilibrium financial market models. \textcolor{black}{Furthermore, we  discuss the passage from ABCEM models to partial differential equations. Here, we especially focus on the perspective of kinetic theory on this limit process.  
In section three, we present the continuous formulations and continuous limits of the Levy-Levy-Solomon and Franke-Westerhoff model. We specifically discuss the impact of different time scalings in the Levy-Levy-Solomon model. Furthermore, we show that a naive numerical discretization of the time continuous Franke-Westerhoff model leads to blow ups. Therefore, we introduce a semi-implicit discretization of the time continuous Franke-Westerhoff model and show that the qualitative output coincides with the original model. In section four we derive the mesoscopic description of the Levy-Levy-Solomon and Franke-Westerhoff model. In addition, we shortly discuss analytical results such as steady states and explicit solutions. We finish this work with a short conclusion. }

\section{Mathematical Perspective on ABCEM Models}

We  introduce the connection of ABCEM models to time continuous dynamical systems. In particular, we state the advantages of time continuous models in comparison to difference equations. 

\label{GenPrice}
As mentioned before, ABCEM models may be interpreted as discretized dynamical systems. Many models in literature neglect the time dependence respectively normalize the time step to one \cite{franke2009validation, harras2011grow, hommes2001financial}. Then in fact, many models are rather formulated by difference equations. In this section, we lay out the connection between difference equations and discretized differential equations. Furthermore, we introduce a rather general model for an \textcolor{black}{out of equilibrium market} which is broadly used in literature. \textcolor{black}{
Finally, we discuss the passage from ABCEM models to partial differential equations. Here, we put an emphasis on kinetic theory and the derivation of mesoscopic descriptions.  \\ 
In genaral one might refer to a market as irrational market if the fixed stock price $S(t_0+k \Delta t)$ at each time step $ t_0 +k\ \Delta t,\ \Delta t>0,\ t_0\geq0,\  k\in\N$ does not clear all buy or sell orders. This corresponds to the situation that the aggregated excess demand is non-zero. Excess demand $D$ can be generally defined as the sum of agents' demand subtracted by the sum of agents' supply. For a detailed discussion of aggregated excess demand we refer to \cite{mantel1974characterization, sonnenschein1972market}.} An early example of an \textcolor{black}{disequilibrium model} is the Beja and Goldman model \cite{beja1980dynamic}. In fact, the Beja and Goldman model can be derived from a rational market model where supply equals demand,
\begin{align}
D(S)\stackrel{!}{=}0. \label{AEDR}
\end{align}
A relaxation of the algebraic demand supply relation \eqref{AEDR} leads to the time continuous ordinary differential equation (ODE)
\begin{align}
\frac{d}{dt} S = \frac{1}{\lambda} D, \label{SODE}
\end{align}
with market depth $\lambda>0$. For detailed discussion of the Beja and Goldman model we refer to the original paper \cite{beja1980dynamic} and recent publications by Trimborn et al. \cite{trimborn2020sabcemm, lax2019disequilibrium}. 
 In many models \cite{day1990bulls, alfarano2008time, lux1995herd, chiarella2002speculative} the price adjustment rule reads,
\begin{align}\label{simple}
S_{k+1} = S_k+ \Delta t\ D_k.
\end{align}
Here, we use $S_k$ as short hand notation for $S_k:= S(t_0+k\ \Delta t)$. 
Often the time step $\Delta t$ is normalized to one such that we are faced with a difference equation. The previous updating rule \eqref{simple}
can be interpreted as an explicit Euler discretization of the ODE \eqref{SODE}, provided that the aggregated excess demand is deterministic.
Stochasticity in the excess demand usually models uncertain external influences such as a news stream or microscopic diversity.
In the case of the Franke-Westerhoff model \cite{franke2012structural}, the aggregated excess demand is stochastic and thus this pricing mechanism cannot be seen as an approximation of an ODE. The pricing rule of the Franke-Westerhoff model can be interpreted as discretized SDE. For details on this specific case, we refer to appendix \ref{appendixModel}. A very general model of an irrational market has been introduced by Trimborn et al. in \cite{trimborn2020sabcemm}. It is given by the following SDE 
 \begin{align}
 dS = F(S, D)\ dt+ G(S,D)\ dW, \label{GenMod}
 \end{align}
 with Wiener process $W$ and arbitrary functions $F$ and $G$.
 \textcolor{black}{The diffusion function $G$ models the influence of the stock price and excess demand on the noise level. The function $F$ models the influence of the sock price change on the excess demand and the stock price itself. } Notice, that \eqref{SODE} is a special case of the model \eqref{GenMod}.
 We use the usual notation for  It\^{o} stochastic differential equations.
 Many market mechanism of ABCEM models are special cases of model \eqref{GenMod}, for example the models presented in \cite{day1990bulls, alfarano2008time, lux1995herd, chiarella2002speculative, chiarella2005dynamic, chiarella2006asset, chiarella2007heterogeneous, challet2001stylized, zhou2007self, andersen2003game, harras2011grow, sornette2006importance, kaizoji2002dynamics, palmer1994artificial, bouchaud1998langevin, cont2000herd, cross2005threshold, cross2007stylized, cross2006mean, dieci2006market, farmer2002price, lux1999scaling, lux2000volatility, de2005heterogeneity}.  \\
The simplest discretization of such an SDE \eqref{DisGenMod} is the Euler-Maruyama method.
 Applying the Euler-Maruyama method to equation \eqref{GenMod}, we obtain:
  \begin{align}
 S_{k+1} = S_k+ \Delta t\ F(S_k, D_k)+ \sqrt{\Delta t}\  G(S_k,D_k)\ \eta,\quad \eta\sim \mathcal{N}(0,1). \label{DisGenMod}
 \end{align}
 In the case of a fully deterministic model, the numerical scheme \eqref{DisGenMod} is identical to the standard Euler method. \\
From a mathematical perspective, we stress that more sophisticated numerical methods for equation \eqref{GenMod} exist, which may improve the quality of approximation remarkably as discussed by Hairer and
 Wanner  \cite{hairer2008solving, wanner1991solving}.
In particular for the case of stiff SDEs or ODEs, one should \textcolor{black}{ consider} implicit solvers to prevent stability problems.
 \textcolor{black}{We refer to the classical references by Kloeden and Wanner  \cite{kloeden2012numerical, wanner1996solving}  for a detailed introduction in the topic of stiff differential equations.  }
 \\ \\
In the ABCEM literature, most models rely on the explicit Euler (in case of deterministic dynamics) or Euler-Maruyama (in case of stochastic dynamics) discretizations.
Often, the numerical approximation is rescaled and fixed such that the time step is set to one.
Hence, in ABCEM literature, we are rather faced with difference equations of the following type
   \begin{align}\label{GenModApp}
 S_{k+1} = S_k+  \bar{F}(S_k, D_k)+ \bar{G}(S_k,D_k)\ \eta,
 \end{align}
than with differential equations. The model \eqref{GenModApp} represents a discretized version of the model \eqref{GenMod} with discretizations $\bar{F}, \bar{G}$ of functions $F, G$.  Finally, we would like to stress that a time continuous model is not only advantageous from a numerical perspective but enables the user to simulate the model on differently coarse time levels by simply adapting the time step in the numerical scheme.

\subsection{Connection to Partial Differential Equations}
\textcolor{black}{
A further advantage of time continuous ODE or SDE models is the possibility to pass to a mesoscopic PDE model. 
In comparison to the microscopic model one looses information, because one does not track each agent individually.  On the mesoscopic level one rather considers a probabilistic interpretation of the agent, since the solution of the PDE is the probability density function of the microscopic law. Thus, instead to consider the wealth of each agent one only knows the probability of an agent to have a certain wealth. 
It is of immanent importance that the mass  is a conserved quantity of the PDE model, otherwise the solution could not be identified as a probability density function. 
The advantage of this probabilistic description is the possibility to analyze the PDE with several mathematical tools. For example, it may be of interest to study the long time dynamics of the system, compute steady state distribution or to derive rates of convergence. In addition, it is possible to derive a macroscopic description out of the PDE model. The unknowns of  a macroscopic model are moments of the probability density  function, and therefore such models are also known as moment models. Moment models can be directly derived by testing the weak form of the mesoscopic PDE model. In many cases such moment models can not be closed and one is faced with the so called closure problem \cite{levermore1996moment}. The general advantage of a moment model is the fast computation of aggregated quantities, which are in the context of ABCEM models e.g. the average stock price or the average wealth. In order to present a simple example we consider $N$ financial agents with wealth $w_i\geq 0,\ i=1,...,N$. We may assume that the wealth of all agents are invested in a bond with fix interest rate $r>0$. 
Thus the wealth dynamic of the i-th agent reads,
\begin{align}
\frac{d}{dt} w_i(t)= r\ w_i(t),\quad w_i(0)=w_i^0.\label{wealthODE}
\end{align}
As we see, is the dynamic identical for each agent, except the initial condition $w_i^0$. The mesoscopic description of an ODE of the type \eqref{wealthODE} is known as Liouville equation. 
For a sufficient number of agents and well posed initial data it is straightforward to interpret the initial wealth as a probability function $f_0(w),\ w\geq 0$ which is transported by the following Liouville type PDE,
\begin{align}
\partial_t f(t,w)+\partial_w( r\ w\  f(t,w) ) =0,\quad f(0,w)=f_0(w).\label{wealthPDE}
\end{align}
Notice that \eqref{wealthPDE} is a simple transport equation and \eqref{wealthODE} is the characteristic equation of the PDE \eqref{wealthPDE}. A simple computation reveals that the  \textcolor{black} {mass $\int f(t,w)\ dw$ is conserved. This is a necessary condition in order to guarantee that a given probability distribution function $f(0,w)$ remains a probability law for any time $t>0$.}
The first moment of $f$,
$$
m(t):=\int\limits_{\R^+} w\ f(t,w)\ dw, 
$$
is the average wealth and the corresponding moment model reads:
$$
\frac{d}{dt} m(t) =  r\ m(t),\quad m(0)=\int\limits_{\R^+} w\ f_0(w)\ dw. 
$$
In this simple case the moment model is closed and the solution can be computed immediately. 
\textcolor{black}{This means that the average wealth (first moment) is independent of any higher moments such as the second moment $\int\limits_{\R^+} w^2\ f(t,w)\ dw, $ and thus can be computed independently of any higher order moment. }
In the case of a stochastic process defined by a SDE, the mesoscopic model is not a transport equation but a PDE of  Fokker-Planck type. 
This passage from SDEs to a mesoscopic description is well understood by the so called Feynman-Kac formula \cite{lapeyre2003introduction}. We assume that all needed regularity assumptions of the SDE \eqref{GenMod} are satisfied. Then the corresponding Fokker-Planck equation of the SDE \eqref{DisGenMod} reads,
\begin{align*}
&\partial_t h(t,s)+ \partial_s(F(s, D)\ h(t,s))= \frac{1}{2} \partial_{ss} (G(s,D)^2\ h(t,s)),\\
&h(0,s)=h_0(s).
\end{align*}
Thus, $h(t,s),\ s\in\R,\ t>0$ denotes the probability density of the stochastic  process defined in \eqref{GenMod}. The dimension of the phase space of $h$ is directly linked to the dimension of the SDE system \eqref{GenMod}. We have stated a simple example in one dimension, in the case of a $d$-dimensional stock price dynamic the phase space of the probability density function becomes $h(t,s),\ s\in\R^d$. 
Hence, a large dimensional dynamical system can be translated into a highly dimensional Fokker-Planck equation. The drawback of a large dimensional PDE are high computational costs and thus for large dimensions there is no computational benefit in comparison to the original model. }

\paragraph{Kinetic Theory}
Kinetic theory describes gases as a large system of interacting atoms. The theory dates back to early works of Daniel Bernoulli or James Maxwell. The core idea of kinetic theory is to consider a mesoscopic description of large particle systems. The previously presented approaches can be seen as special cases of particle systems which can be described by kinetic theory. The Liouville equation is derived from Newton's equation in the case of no particle interactions. Thus, the particles only differ in their initial configuration, as in our example \eqref{wealthODE}.
The Feynman-Kac approach can be applied to arbitrary dimensional SDE systems, but the phase space of the resulting Fokker-Planck equation may become too large and unfeasible. 
There exists two famous kinetic limits which avoid the problem of dimension and describe the passage from microscopic dynamics to a mesoscopic description, namely the Boltzmann-Grad limit and the mean field limit. Both limits have been analyzed profoundly and are still subject to current research \cite{bolley2011stochastic, dobrushin1979vlasov, lanford1975time, marklof2011boltzmann}. In the Boltzmann-Grad limit one considers binary interactions of particles and the resulting mesoscopic model is the Boltzmann equation of gas dynamics originallly proposed by Boltzmann in 1871 \cite{cercignani1988boltzmann, boltzmann2012lectures}.
In the mean field setting one rather considers interactoins of one particle with the field of all other particles. The classical example of a mean field equation is the Vlasov-Poisson equation, derived by Vlasov in 1938 \cite{vlasov1938oscillation}. The Vlasov-Poisson equation models charged particles in a plasma. \\ \\
These kinetic limits need to be understood as an approximation of the microscopic particle dynamics and are only valid in the regime of infinitely many particles \cite{golse2016dynamics}.  
These limits can be applied provided that a certain symmetry structure of microscopic dynamics is given. The reduction in computational costs is enormous since the phase space of the kinetic model coincides with the dimension of the dynamics of one particle. There is also a link from kinetic equations to macroscopic models such as the Euler or Navier-Stokes equations. These connections can be obtained by a moment model of the Boltzmann equation equipped with the corresponding asymptotic limit \cite{de1989incompressible, grad1964asymptotic, saint2009hydrodynamic}. \\ \\
In the past decade kinetic theory has been applied to a variety of biological and socio-economic systems \cite{bellomo2017active, bellomo2019active, pareschi2013interacting, naldi2010mathematical}. In this context we rather speak of agents instead of physical particles. On the other hand the modeling of life-science and socio-economic phenomena by agent-based systems have increased rapidly \cite{gilbert2019agent, bonabeau2002agent, railsback2019agent}. There are several examples of kinetic models derived from existing agent-based models \cite{albi2014kinetic, albi2017recent, trimborn2017portfolio, trimborn2018mean, pareschi2013interacting}. We aim to give toy example of microscopic wealth interactions which can be conveniently studied by kinetic theory. As before we consider $N$ financial agents equipped with wealth $w_i\geq 0$ and dynamic,
\begin{align}\label{simpODE}
\frac{d}{dt} w_i =  \frac{1}{N} \sum\limits_{k=1}^N w_k-w_i,\quad w_i(0)= w_i^0.
\end{align} 
This dynamic has been previously studied by Bouchaud and M\'ezard \cite{bouchaud2000wealth} and Pareschi \cite{pareschi2013interacting}. In order to derive the mean field limit of our microscopic ODE system \eqref{simpODE} it is convenient to introduce the following tool. 
\begin{definition}
Given a vector $\bs{x}:=(x_1,...,x_N)^T\in \mathbb{R}^{dN},\ x_i:=(x_i^1,...,x_i^d)\in\R^d,\  i=1,...,N$ then the empirical measure (or atomic probability measure) $\mu^N_{\bs{x}}$ is defined as,
\[
\mu^N_{\bs{x}}(x^1,...,x^d)=\frac{1}{N} \sum\limits_{i=1}^N \delta(x^1-x_i^1)\cdot...\cdot \delta(x^d-x_i^d).
\]
\end{definition}
The empirical measure enables to connect the solution of the microscopic ODE system \eqref{simpODE} to the solution of the corresponding mean field equation. The Picard-Lindelöf theorem ensures existence of a unique solution of \eqref{simpODE}, which we denote as $\boldsymbol{w}(t)=(w_1,...,w_N)'\in(\R^+)^N$. Then we test the corresponding empirical measure $\mu_{\boldsymbol{w}}^N(t,w),\ w\in\R^+$ with the test function $\phi(w)$ and compute,
\begin{align*}
\frac{d}{dt} \int\limits_{\R^+} \phi(w)\ \mu_{\boldsymbol{w}}^N(t,w)\ dw &= \frac{1}{N} \sum\limits_{k=1}^{N}  \frac{d}{dt}\phi(w_k(t))=  \frac{1}{N} \sum\limits_{k=1}^{N}  \partial_w\ \phi(w_k(t))\ \frac{d}{dt} w_k(t)\\
& = \frac{1}{N} \sum\limits_{k=1}^{N}  \partial_w\ \phi(w_k(t))\  \frac{1}{N} \sum\limits_{j=1}^N w_j(t)-w_k(t)\\
&= \frac{1}{N} \sum\limits_{k=1}^{N}  \partial_w\ \phi(w_k(t))\  \int\limits_{\R^+}  w' \ \mu^N_{\boldsymbol{w}}(t,w')\ dw'-w_k(t)\\
& = \int\limits_{\R^+} \partial_w \phi(w)\ \left(\int\limits_{\R^+}  w' \ \mu^N_{\boldsymbol{w}}(t,w')\ dw'-w\right)\ \mu^N_{\boldsymbol{w}}(t,w)\ dw.
\end{align*} 
\textcolor{black}{In this probabilistic setting, a test function can be understood as the observable quantity of a random variable, e.g. in the case of the identity $\phi(w)= w $ we obtain the expected value of the probability law. Note that the derivative of the empirical measure w.r.t $w$ is not directly applicable. After multiplication of the empirical measure against a test function $\phi(w)$ and integration, the derivative w.r.t. $w$ may be moved onto the test function via partial integration. This standard method in variational calculus allows us to link the point-wise declared empirical measure to an integral equation. Hence,  the empirical measure $\mu^N_{\boldsymbol{w}}$ is a weak solution of the mean field PDE,}
\begin{align}\label{MFToy}
\partial_t f(t,w)+ \partial_w\left( \left[ \int\limits_{\R^+} w'\ f(t,w')\ dw'-w  \right]\ f(t,w)   \right) = 0,
\end{align}
with initial conditions $f(0,w)= \frac{1}{N} \sum\limits_{k=1}^N\delta(w-w_k^0)$. The mean field PDE \eqref{MFToy} is an integro-differential equation which models the behavior of the system \eqref{simpODE}
in the limit of infinitely many agents. The convergence of the microscopic system to the mean field equation can be proven rigorously, following the theory of Dobrushin or Neunzert \cite{dobrushin1979vlasov, neunzert1977vlasov}. The rigorous derivation of the mean field equation is performed in section \ref{PropSection} for the Levy-Levy-Solomon model.

\section{Continuum Limit and Robust Formulation}
\label{examples}
In this section we discuss the robust formulation of ABCEM models with help of the Levy-Levy-Solomon model and the Franke-Westerhoff model.
As pointed out in the introduction are both models structurally very different. The Levy-Levy-Solomon model considers $N$ agents whereas the Franke-Westerhoff model can be regarded as
a two agent model. \\
We introduce possible continuous formulations of the Levy-Levy-Solomon and Franke-Westerhoff model. Furthermore, we show the impact of different continuous formulations on the wealth evolution in the Levy-Levy-Solomon model. In addition, we discuss the impact of different numerical methods on the model behavior of the Franke-Westerhoff model. \\ \\
All presented results have been generated with the SABCEMM simulator which is freely available on GitHub \cite{SABCEMMgithub}.
For the used pseudo random number generators used in obtaining the results, please confer to table \ref{tab:RNGTable}.
%Furthermore, the simulation data is published \cite{SabnumData} such that the reader can reproduce the presented results.

\subsection{Continuum Limit}
\label{CLimit}
In this section, we introduce time continuous versions of the Levy-Levy-Solomon and Franke-Westerhoff model. As usual in the ABCEM literature, the models are originally formulated as difference equations. 
Obviously, such a continuum limit is not uniquely defined and, in the case of the Levy-Levy-Solomon model, we discuss several different time discretizations in the agent dynamic. Thus, we focus on the impact of different time discretizations in the Levy-Levy-Solomon model and derive the continuum limit of the Franke-Weserhoff model which exhibits stability problems in the case of an explicit Euler discretization. A detailed discussion of the continuum version of the Levy-Levy-Solomon and Franke-Westerhoff model can be found in the appendix \ref{appendixModel}.

\begin{figure}[ht]
\centering
\begin{subfigure}{0.49\linewidth}
\includegraphics[width=\linewidth]{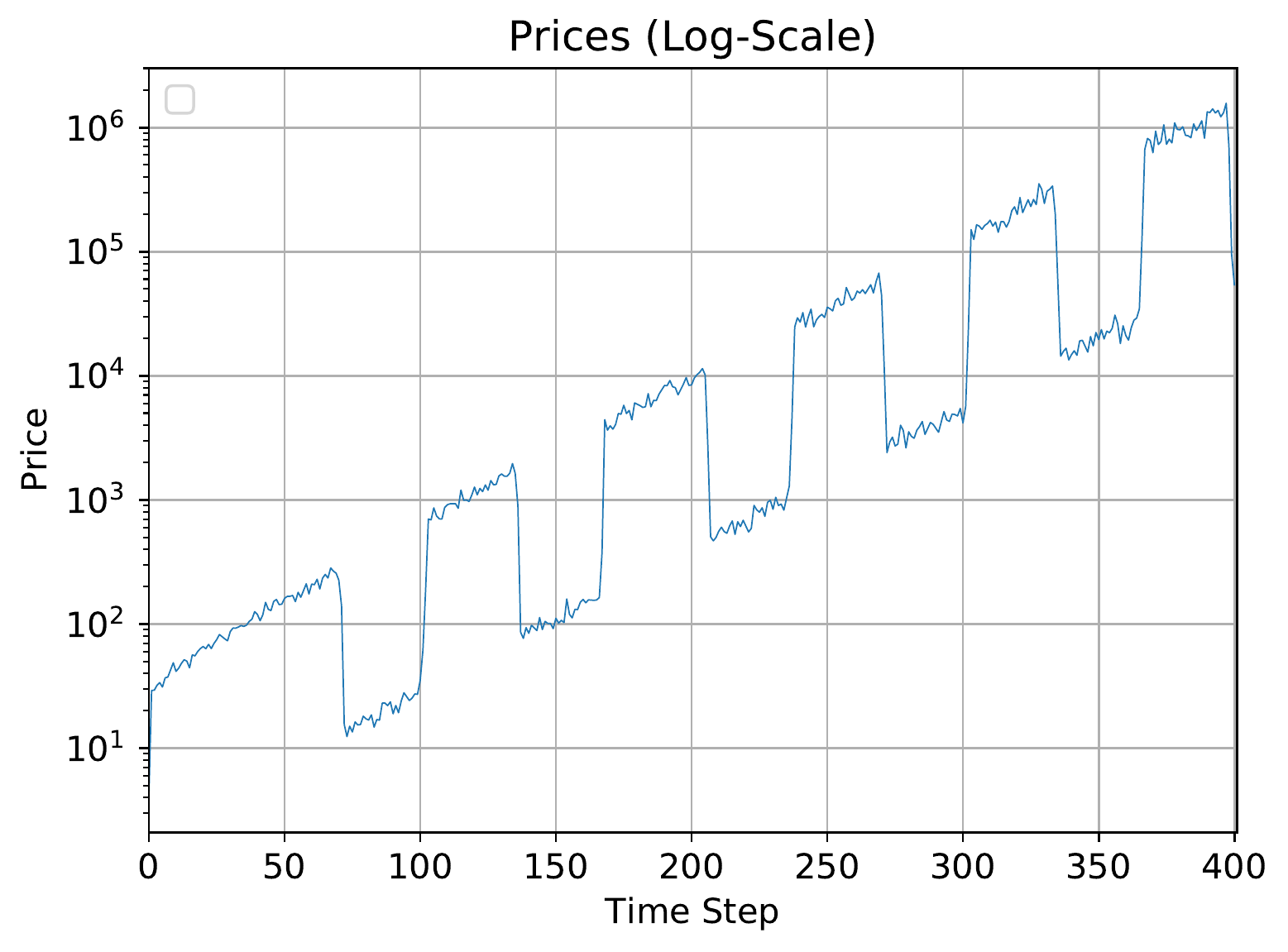}
\caption{$\Delta t=0.5$}
\end{subfigure}
\begin{subfigure}{0.49\linewidth}
\includegraphics[width=\linewidth]{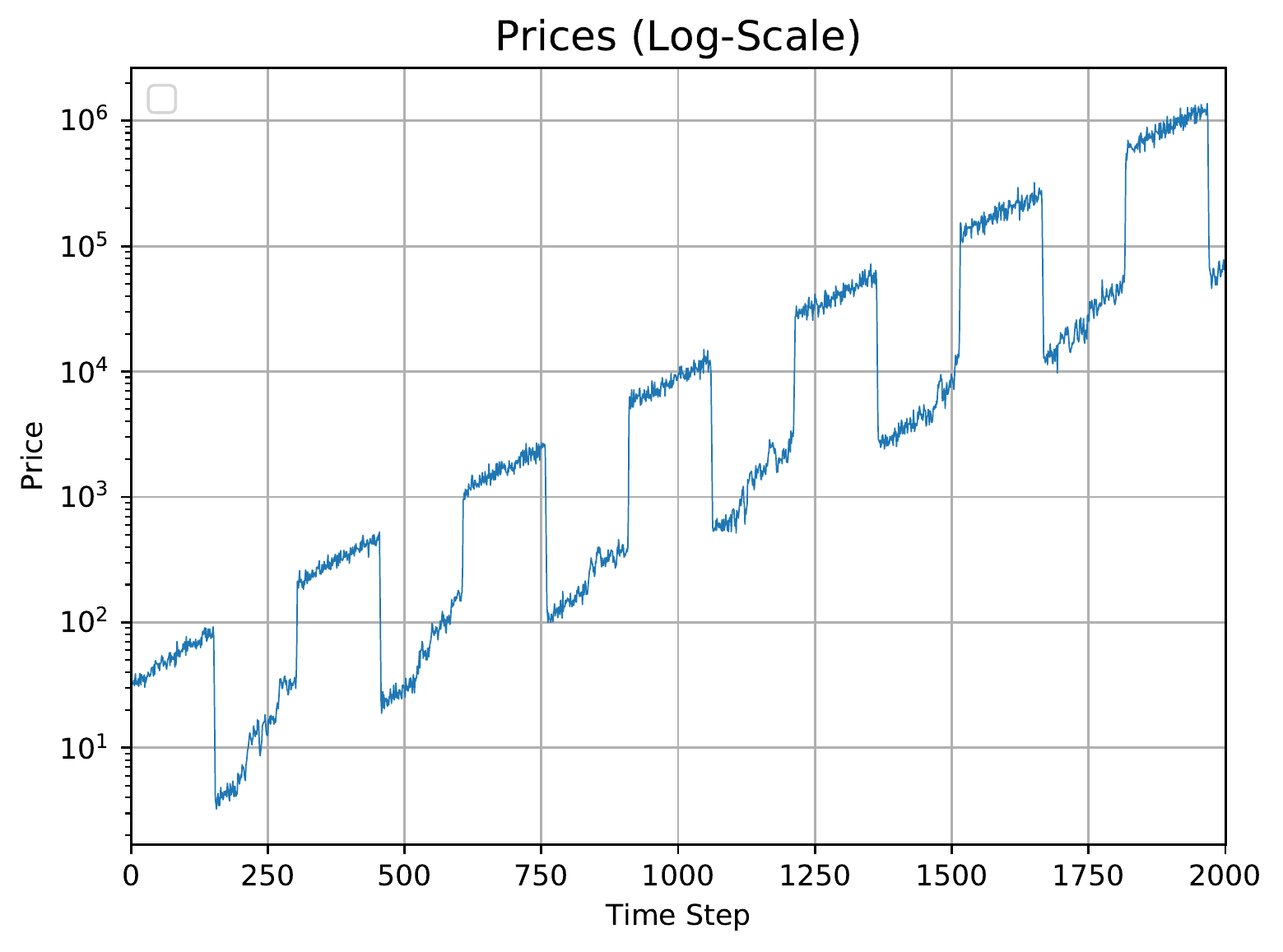}
\caption{$\Delta t=0.1$}
\end{subfigure}
\begin{subfigure}{0.49\linewidth}
\includegraphics[width=\linewidth]{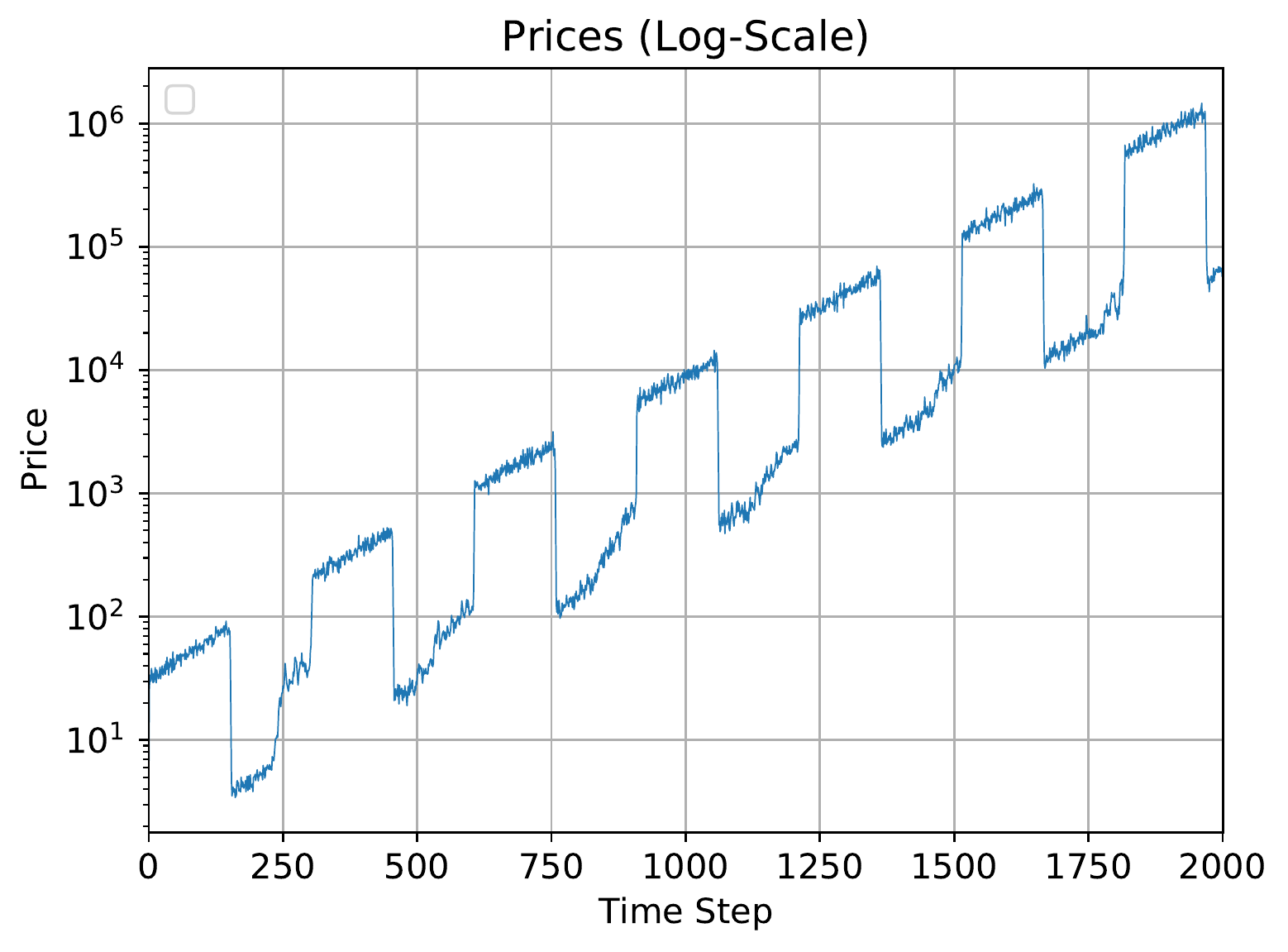}
\caption{$\Delta t=0.05$}
\end{subfigure}
\begin{subfigure}{0.49\linewidth}
\includegraphics[width=\linewidth]{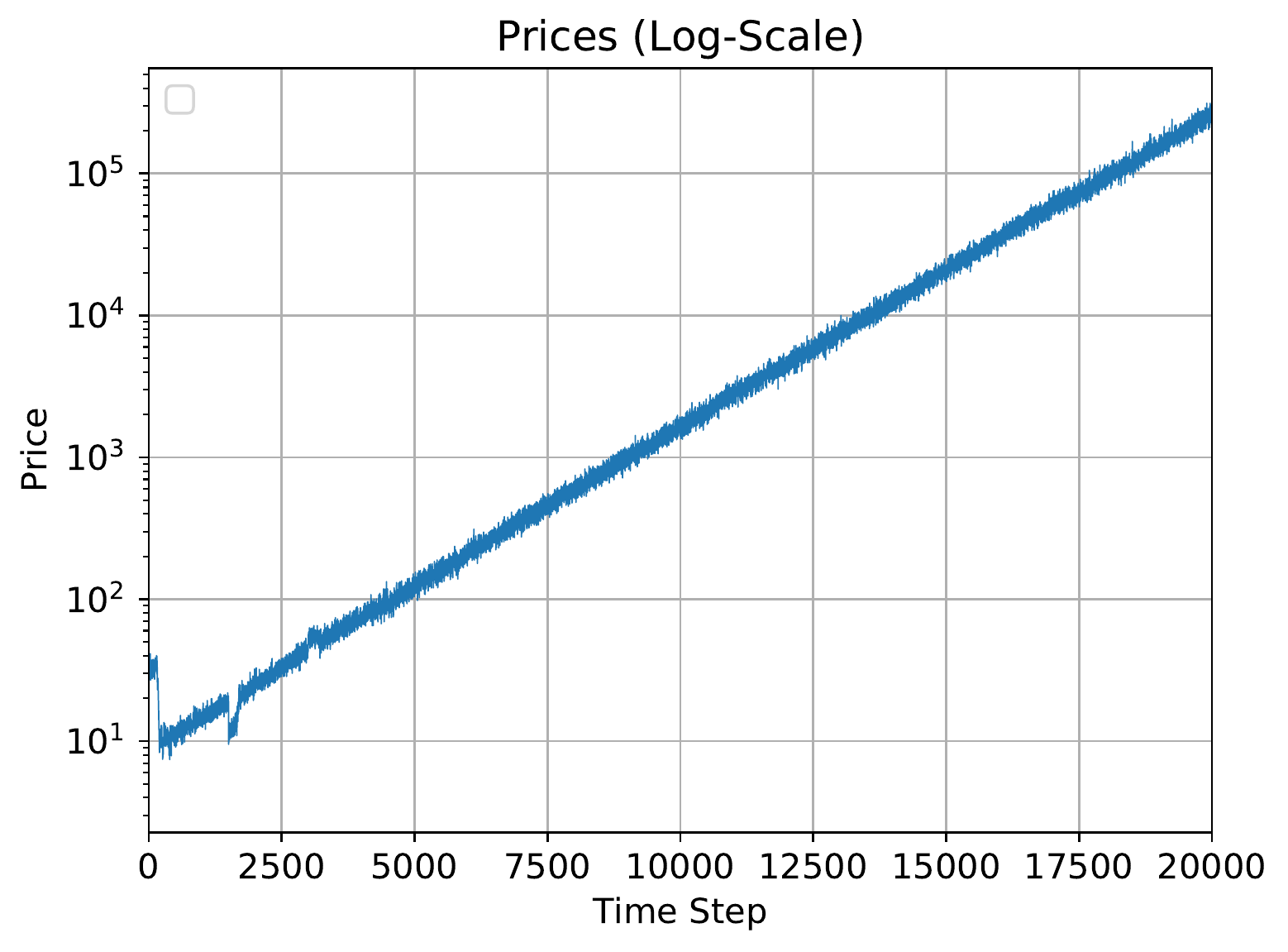}
\caption{$\Delta t=0.01$}
\end{subfigure}
\caption{Simulations of time continuous Levy-Levy-Solomon model with scaled memory variable and different time discretizations. Further parameters as defined in table \ref{LLS-basic} with $\sigma_{\gamma}=0.2$.}
\label{LLS-DeltaT}
\end{figure}

\paragraph{The Levy-Levy-Solomon Model }
\label{TimeDiscretization}
In the following, we introduce a time scale respectively the time step $\Delta t>0$ in order to perform the continuum limit in a second step.
Interpreting the original Levy-Levy-Solomon model as the result of an explicit Euler discretization where $\Delta t$ is set to 1, a general time-discretized version of the wealth evolution is given by:

\begin{align}
w(t+\Delta t) = w(t)+ \Delta t \left[ (1-\gamma(t))\ r +  \gamma(t)\ \frac{\frac{S(t+\Delta t)- S(t)}{\Delta t}+Z(t)}{ S(t)} \right]\ w(t).\label{LLScontinuous}
 \end{align}
For a proper definition of all parameters and functions we refer to the appendix \ref{appendixModel}.
Notice that the bond return $r$ and the stock return $\frac{\frac{S(t+\Delta t)-S(t)}{\Delta t} + Z(t)}{ S(t)}$ are rates and thus scale in time.
Equation \eqref{LLScontinuous} represents an explicit Euler discretization of the ODE,
\begin{align*}
\frac{d}{dt} w(t) = \left[ (1-\gamma(t))\ r + \gamma(t)\  \frac{\frac{d}{dt} S(t)+Z(t)}{S(t)} \right] \ w(t),
\end{align*}
where the time differential $\frac{d}{dt} S(t)$ is approximated by the forward difference quotient \newline $\frac{S(t+\Delta t)-S(t)}{\Delta t}$.
In order to study the time continuous version of the model, we need to properly define the time scaling of the investor.
We want to emphasize that several reasonable time scales exist.
First, we study the case in which the number of recent timesteps an agent $i$ considers in their decision, denoted by variable $m_i$, scales with time, which means: $\bar{m}_i:=\lfloor \frac{m_i}{\Delta t}\rfloor$.
The results for different time steps using an explicit Euler discretization can be seen in figure \ref{LLS-DeltaT}.

As pointed out in \cite{beikirch2018simulation}, approximately $90\%$ of the optimal investment decisions $\gamma_i$ in the original model are located at the boundaries of the interval in $[0.01,0.99]$. Interestingly, in the case of the previously introduced time scaling of the memory variable, this model characteristic changes. For sufficiently small $\Delta t$ the optimal investment decisions ($\gamma_i$ ) are all located in $(0.01,0.99)$ and not at the boundaries.  
This can be explained by the very small optimization horizon and the smoothing effect of a large return history.
For $\Delta t = 0.1$ the percentage of extreme decisions reduces to $72\%$ and for $\Delta t=0.01$ all optimal investment decisions are located in the interior.
Note that these statements are based on the average of the results of over $100$ runs.

\begin{figure}[ht]
\centering
\begin{subfigure}{0.49\linewidth}
\includegraphics[width=\linewidth]{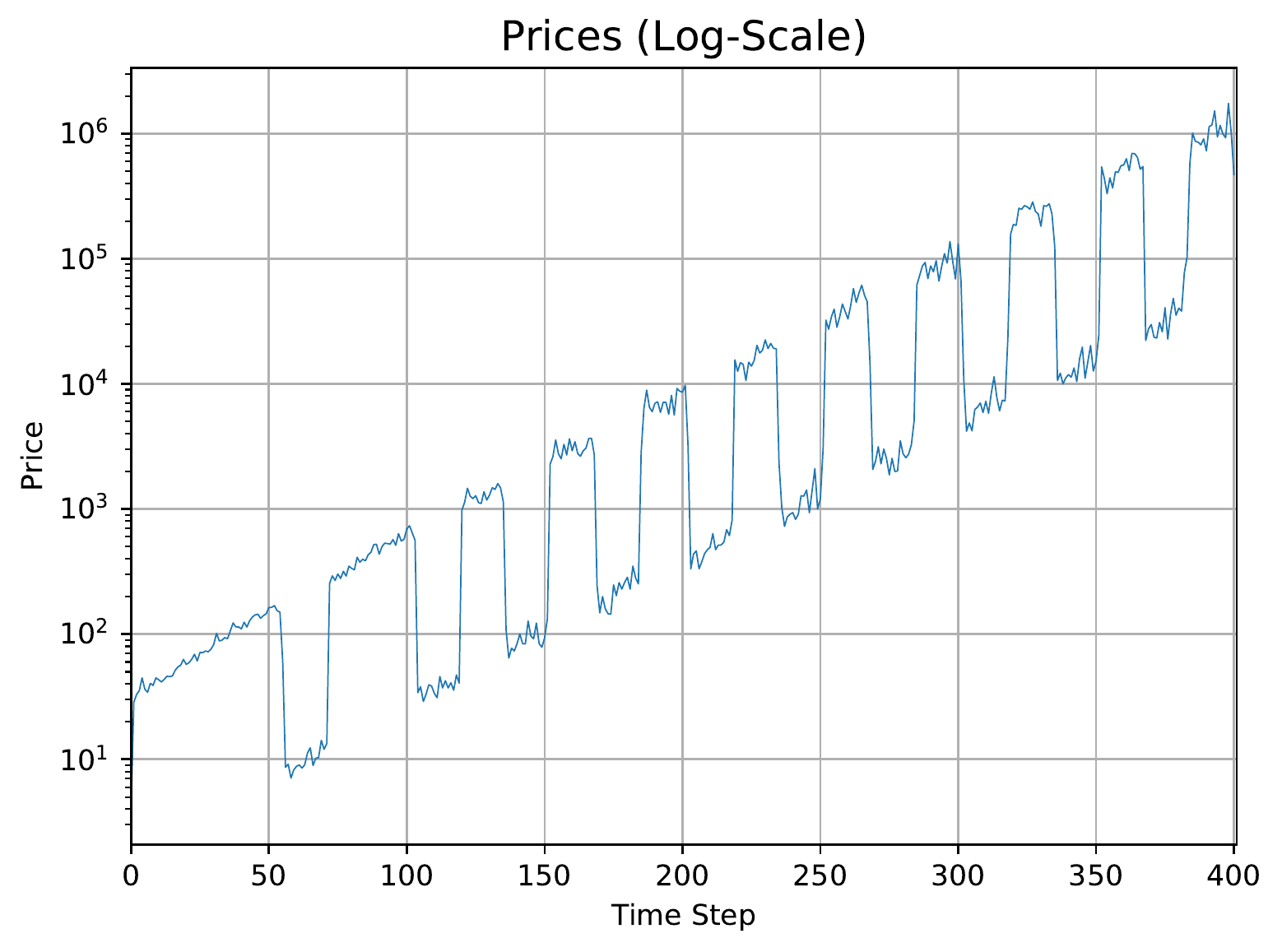}
\caption{$\Delta t=0.5$}
\end{subfigure}
\begin{subfigure}{0.49\linewidth}
\includegraphics[width=\linewidth]{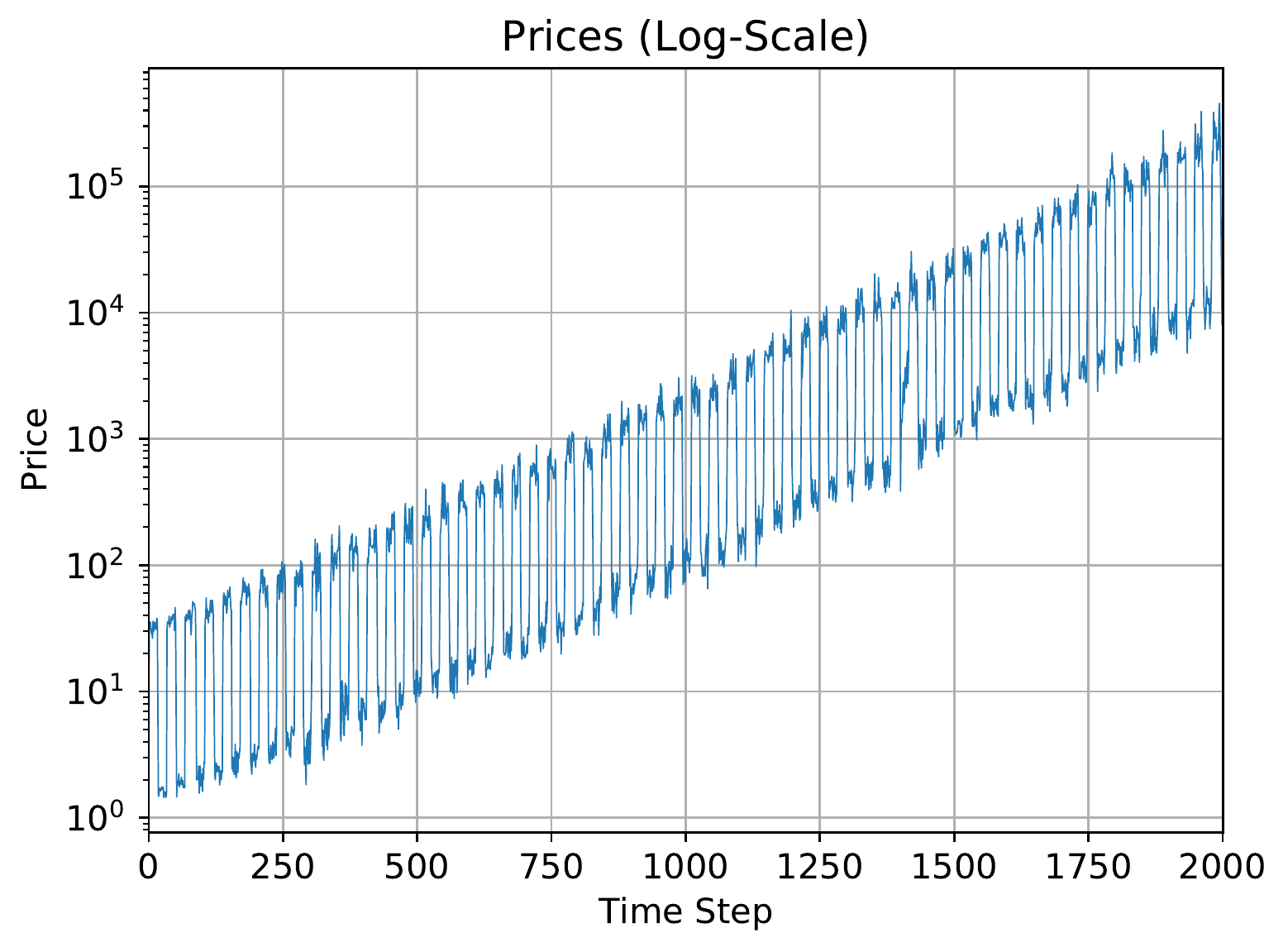}
\caption{$\Delta t=0.1$}
\end{subfigure}
\begin{subfigure}{0.49\linewidth}
\includegraphics[width=\linewidth]{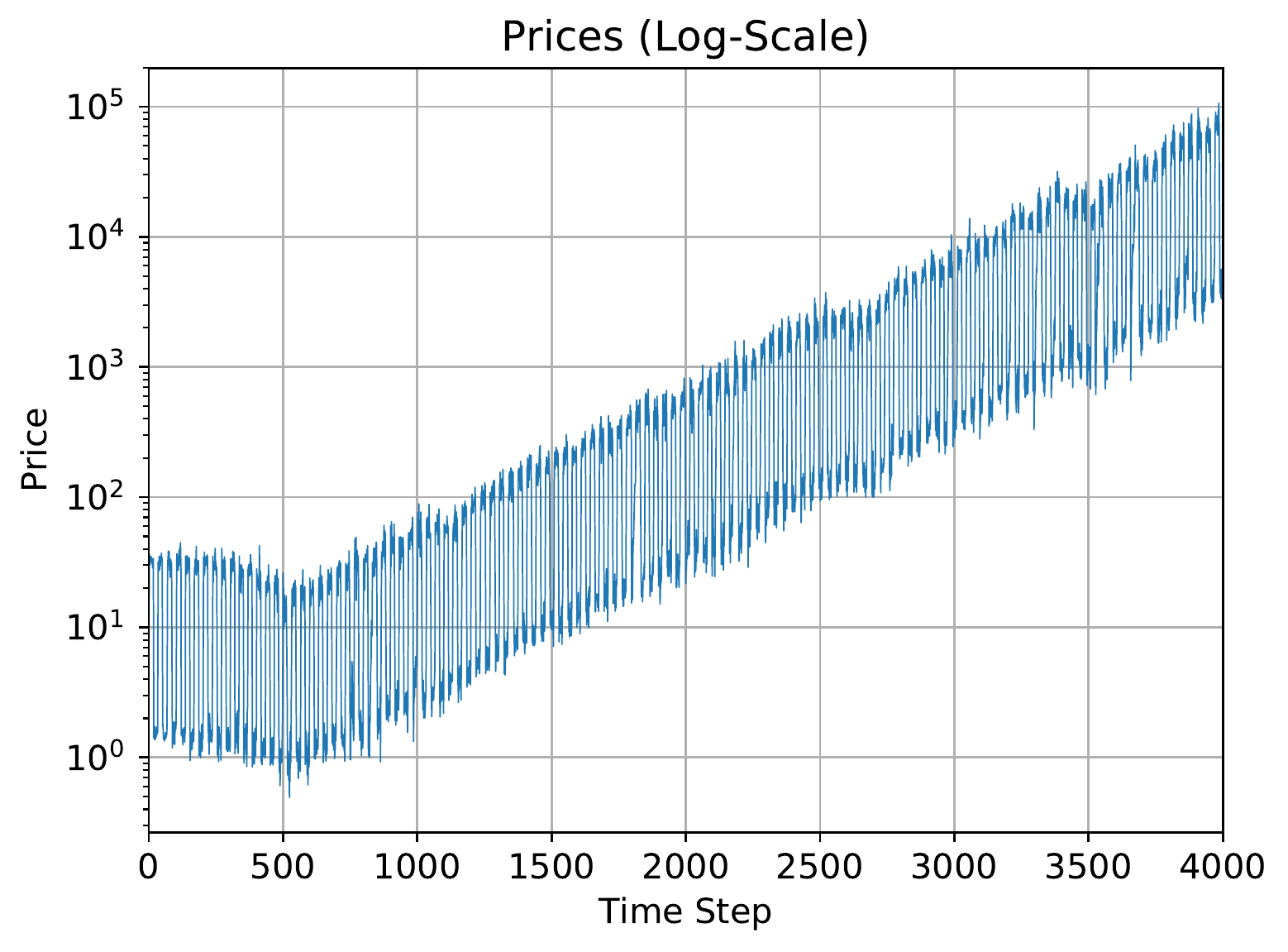}
\caption{$\Delta t=0.05$}
\end{subfigure}
\begin{subfigure}{0.49\linewidth}
\includegraphics[width=\linewidth]{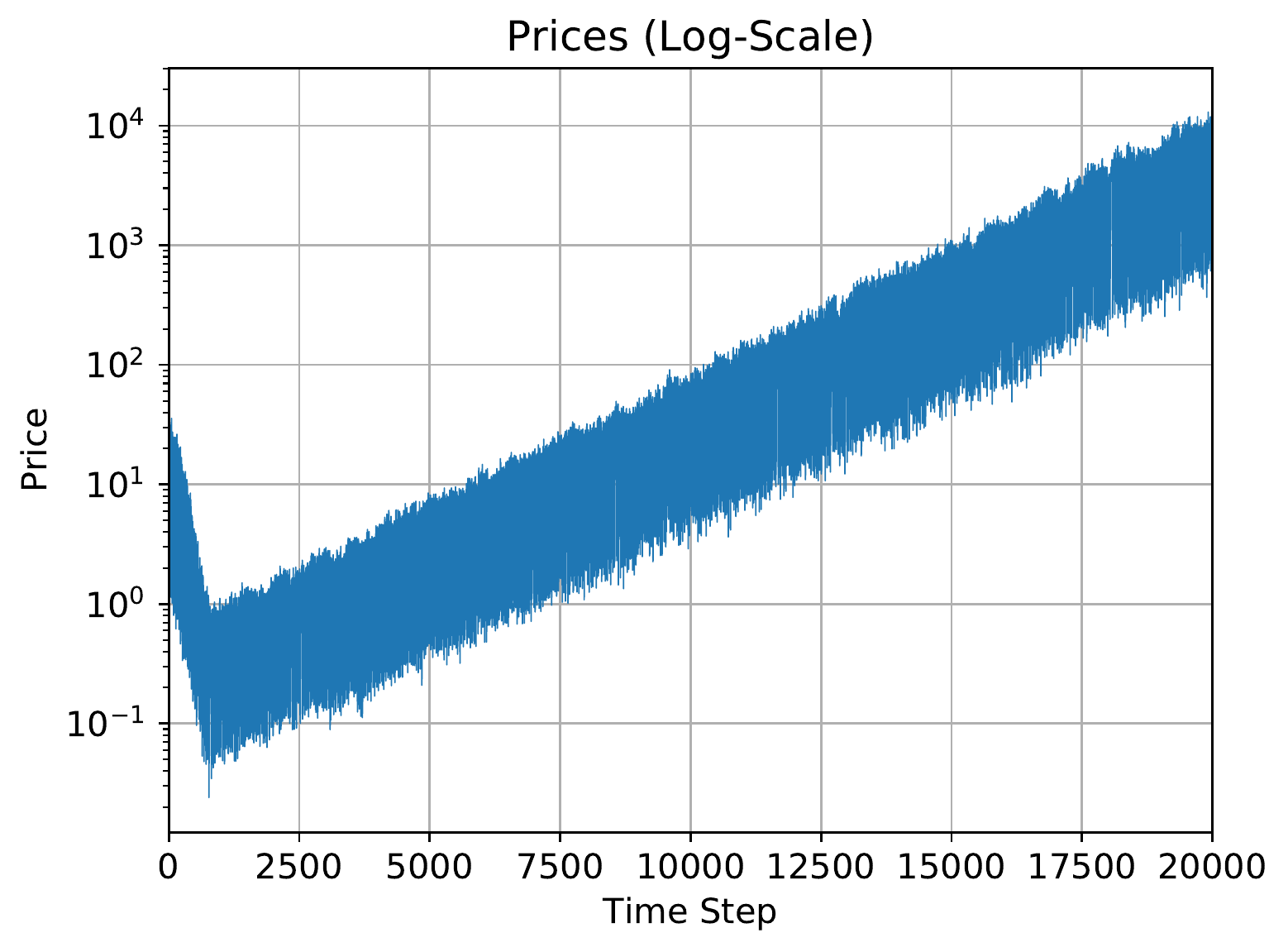}
\caption{$\Delta t=0.01$}
\end{subfigure}
\caption{Simulations of time continuous Levy-Levy-Solomon model with fixed memory variable and different time discretizations. Further parameters as defined in table \ref{LLS-basic} with $\sigma_{\gamma}=0.2$.}
\label{LLS-DeltaTfixedM}
\end{figure}

Alternatively, we may assume that the investor's memory does not scale with time, i.e. the number of time steps which corresponds to the agents' memory is always constant.
 Simulations show that using a non-scaling memory, i.e. with a memory of a fixed number of time steps, oscillating prices for an explicit Euler discretization can be observed for all chosen timesteps (see \cref{LLS-DeltaTfixedM}).
Averaging over $100$ runs also indicates that the percentage of extreme decisions remain approximately around $90\%$ for any chosen time discretization.
The possibility to study further scales of the Levy-Levy-Solomon model is left for future research.

\paragraph{The Franke-Westerhoff Model}
Analog to the Levy-Levy-Solomon model, we interpret the Franke-Westerhoff Model as an explicit Euler-type discretization of a a system of ordinary and stochastic differential equations with $\Delta t = 1$. Under this assumption, we introduce the following rescaled version of agents' dynamics in the Franke-Westerhoff Model as a first step towards a time-continuous model ($\Delta t>0$),
\begin{align}
\begin{split}\label{threeFW}
&n^f(t+\Delta t) = n^f(t) + \Delta t\ n^c(t) \pi^{cf}(a(t,P,n^f,n^c)) - \Delta t\ n^f(t) \pi^{fc}(a(t,P,n^f,n^c)),\\
&n^c(t+\Delta t) = n^c(t) + \Delta t\ n^f(t) \pi^{fc}(a(t,P,n^f,n^c)) - \Delta t\ n^c(t) \pi^{cf}(a(t,P,n^f,n^c)).
\end{split}
\end{align}
For a detailed definition of the model we refer to the appendix \ref{appendixModel}.
The corresponding ordinary differential equations (ODE) of the dynamics \eqref{threeFW} are given by,
\begin{align}
\begin{split} \label{agentdynC}
&\frac{d}{dt} n^f(t) =   n^c(t) \pi^{cf}(a(t,P,n^f,n^c)) - n^f(t) \pi^{fc}(a(t,P,n^f,n^c)),\\
&\frac{d}{dt} n^c( t) =  n^f(t) \pi^{fc}(a(t,P,n^f,n^c)) -  n^c(t) \pi^{cf}(a(t,P,n^f,n^c)).
\end{split}
\end{align}
As the stock price dynamics include random terms, we interpret the original model as an Euler-Mayurama discretization. Hence, we obtain the following rescaled stock price dynamics,
\begin{align}
P(t) = P(t-\Delta t) + \mu\ \Delta t\ D^{FW}(t)+ \sqrt{\Delta t}\ \mu\   \textcolor{black}{(n^f(t)\ \sigma_f+ n^c(t)\ \sigma_c)}\ \eta,\quad \eta\sim \mathcal{N}(0,1). \label{oneFW}
\end{align}
For a detailed definition of $D^{FW}$ we refer to appendix \ref{appendixModel}. 
The SDE corresponding to equation \eqref{oneFW} is given as,
\begin{align}
dP = \mu\ D^{FW}(t)\ dt + \mu \textcolor{black}{ (n^f(t)\ \sigma_f+ n^c(t)\ \sigma_c)}\ dW. \label{FW SDE}
\end{align}
The SDE is interpreted in the It\^o sense and the usual notation for SDEs is employed.
Hence, the time continuous Franke-Westerhoff model reads,
\begin{align}
\begin{split} \label{ODEFW}
&dP( t) = \mu\ D^{FW}(t)\  dt+ \mu  \textcolor{black}{(n^f(t)\ \sigma_f+ n^c(t)\ \sigma_c)}\ dW,\\
&\frac{d}{dt} n^f(t) =   n^c(t) \pi^{cf}(a(t,P,n^f,n^c)) - n^f(t) \pi^{fc}(a(t,P,n^f,n^c)),\\
&\frac{d}{dt} n^c( t) =  n^f(t) \pi^{fc}(a(t,P,n^f,n^c)) -  n^c(t) \pi^{cf}(a(t,P,n^f,n^c)).
\end{split}
\end{align}

We derived the ODE-SDE system \eqref{ODEFW} from the original Franke-Westerhoff Model via the rescaled ODE system \eqref{threeFW} and the SDE \eqref{FW SDE}. 
For a detailed introduction to the Franke-Westerhoff model, we refer to the appendix \ref{appendixModel}. In order to clarify the validity of the derived continuum limit, 
we perform numerical tests. \\

\begin{figure}[htb]
    \centering
    \includegraphics[width=\linewidth, height=0.3\linewidth]{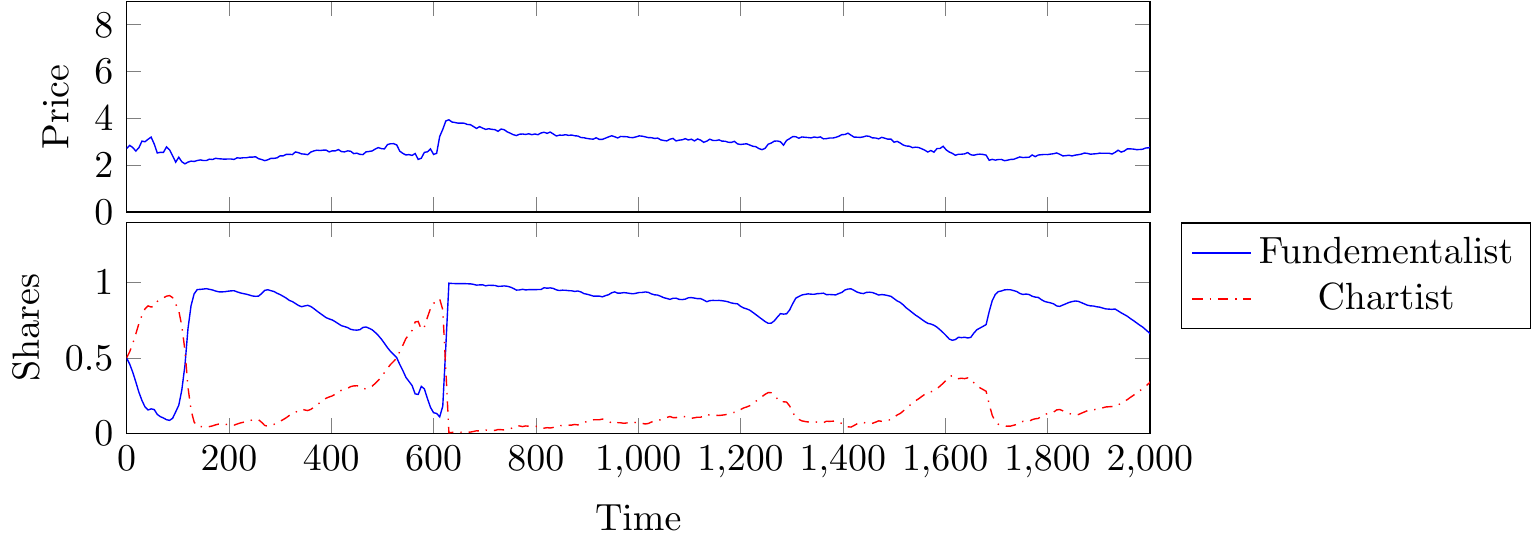}
    \caption{Franke-Westerhoff model with explicit Euler discretization. Parameters as in table \cref{tab:fw_basic}.}
    \label{fig:tpac_dt1_noblowup}
\end{figure}

\begin{figure}[htb]
    \centering
    \includegraphics[width=\linewidth, height=0.3\linewidth]{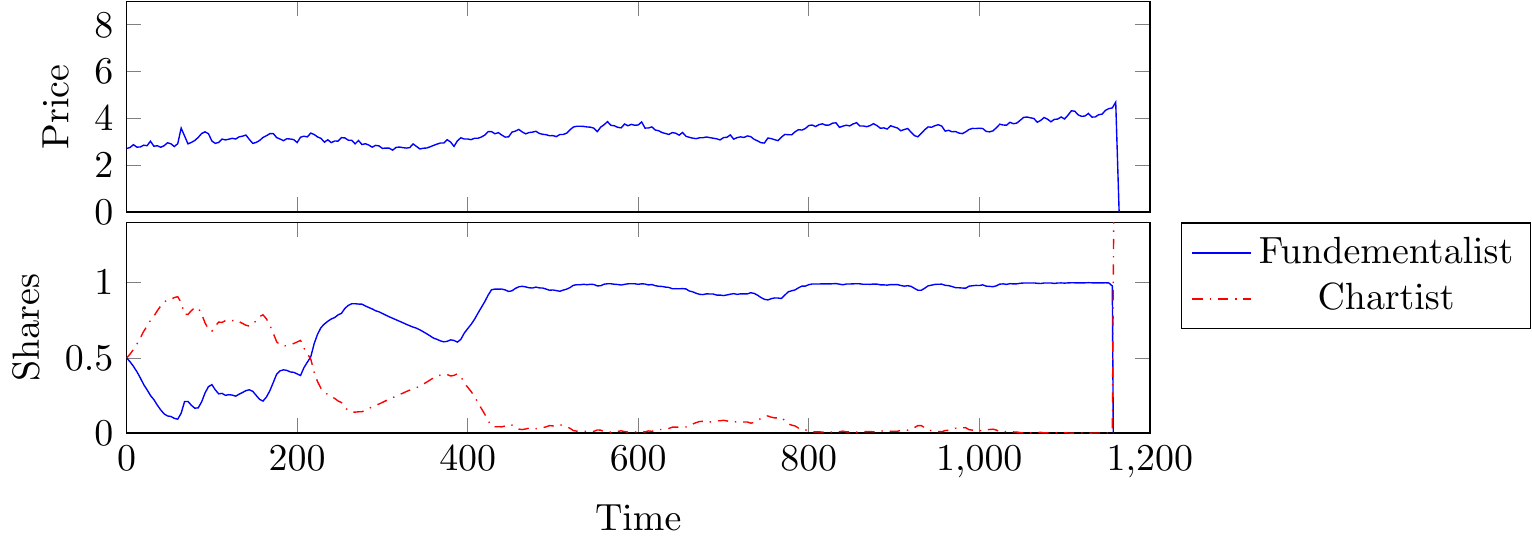}
    \caption{Blow up in the dynamics of the Franke-Westerhoff model with explicit Euler discretization. Parameters as in table \cref{tab:fw_basic} with $\sigma_f=1.15$.}
    \label{fig:tpac_dt1}
\end{figure}

\begin{figure}[htb]
    \centering
    \includegraphics[width=\linewidth, height=0.3\linewidth]{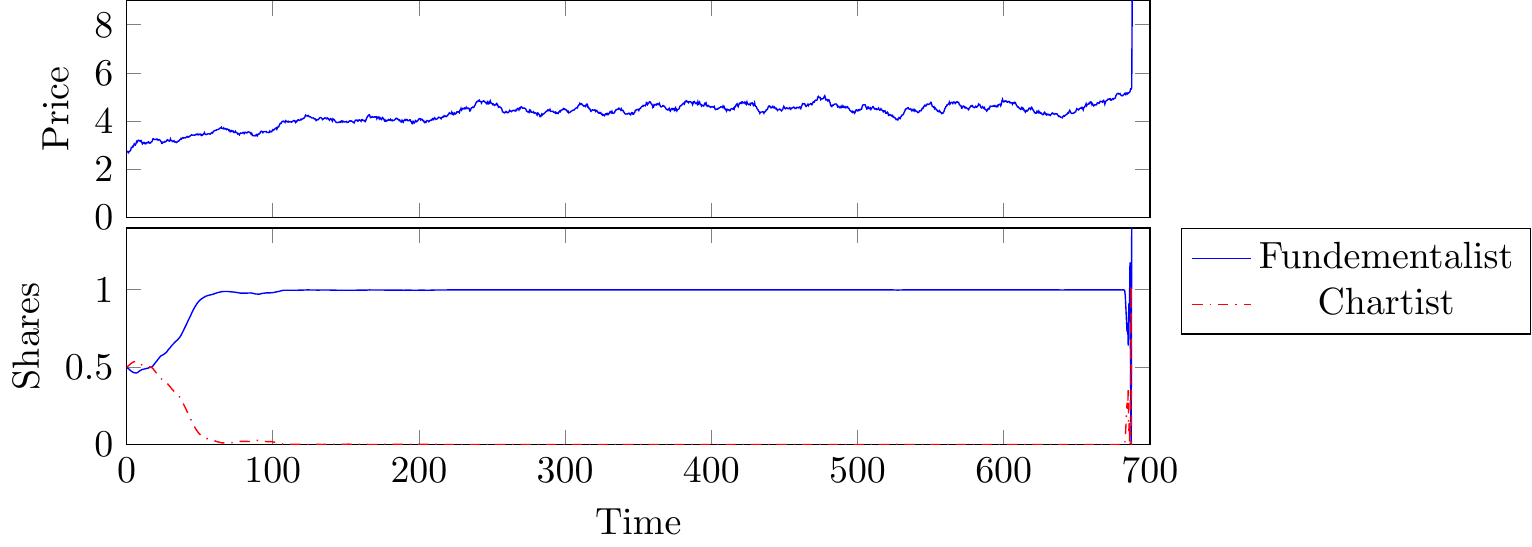}
    \caption{Blow up of the dynamics of the Franke-Westerhoff model with explicit Euler discretization. Parameters as in table \cref{tab:fw_basic} with $\sigma_f=1.15$ and $\Delta t = 0.1$.}
    \label{fig:tpac_dt0-1}
\end{figure}

\begin{figure}[htb]
    \centering
    \includegraphics[width=\linewidth, height=0.3\linewidth]{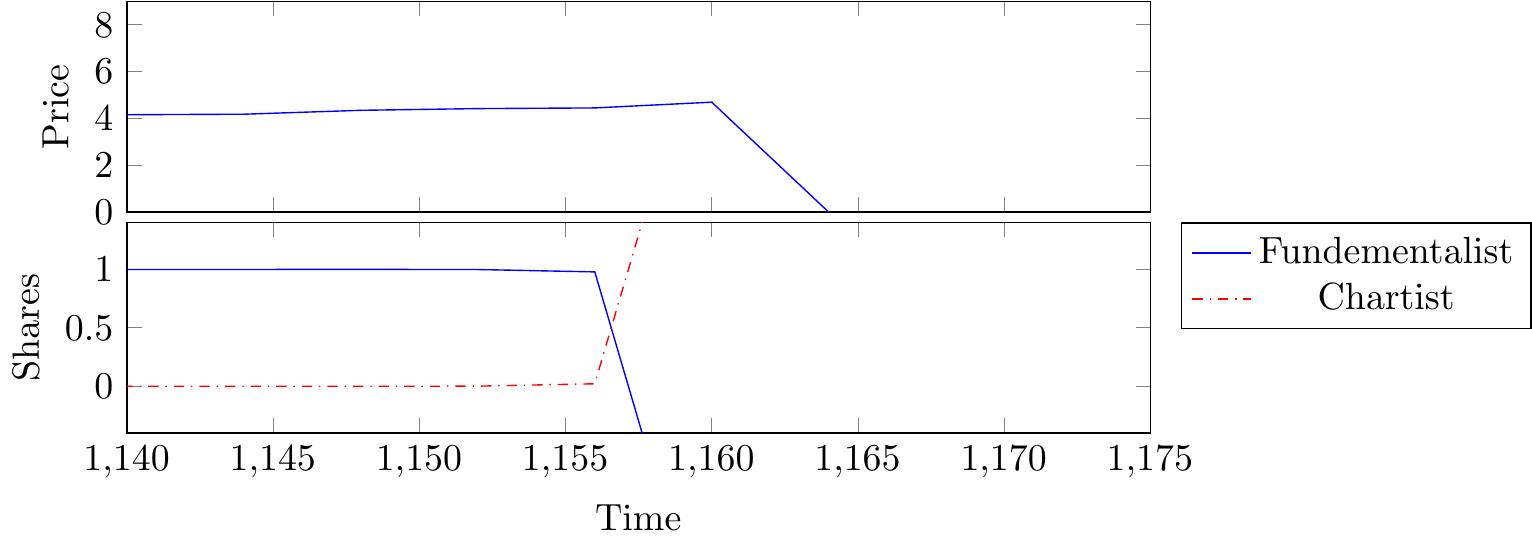}
    \caption{Zoom on the instability of the Franke-Westerhoff model with explicit Euler discretization. See \cref{fig:tpac_dt1} for full plot. Parameters as in table \cref{tab:fw_basic} with $\sigma_f=1.15$.}
    \label{fig:tpac_dt1zoom}
\end{figure}

We first run the model with the parameters presented by Franke and Westerhoff \cite{franke2012structural} and choose the time step $\Delta t$ to be $\Delta t=1$.
The qualitative results are identical to the original model (see \cref{fig:tpac_dt1_noblowup}).
If we change the noise level of the fundamentalist to $\sigma_f = 1.15$, we obtain a blow up of the dynamics (see \cref{fig:tpac_dt1}). By blow up we mean that the numerical solution of our equation tends to infinity at finite time. This is an undesirable model characteristic since a minor change in the model parameters has led to an unfeasible model. \\
This is expected as the only difference is the missing additional constraint \eqref{additional} for the switching probabilities (cf. Remark \ref{addRemark} in appendix \ref{appendixModel}).
\\
%If we change the noise level of the fundamentalist we obtain a blow up of the dynamics (see \cref{fig:tpac_dt1}). This is a undesirable model characteristic since a minor change in the model parameters has led to an unfeasible model.
\\
One might expect that the large time step $\Delta t=1$ may be the reason for the numerical instability. The \cref{fig:tpac_dt0-1} reveals that we still obtain a blow up even for time step $\Delta t= 0.1$.
\\
The reason for the blow up are visible in \cref{fig:tpac_dt1zoom}. The values for $n^f(t)$ and $n^c(t)$ leave the interval of $[0,1]$ while preserving the relation $n^f(t) + n^c(t) = 1$. Subsequently this leads  to a failure in the price calculation.
\\
It appears Franke and Westerhoff have been aware of this model behavior since they have stated the following additional constraint in \cite{franke2011estimation, franke2012structural}, 
\begin{align}
\begin{split}\label{additional}
\pi^{cf}(a(t_{k-1})) &= \min\{1, \nu  \exp(a(t_{k-1}))\},  \\
\pi^{fc}(a(t_{k-1}) &= \min\{ 1, \nu  \exp(- a(t_{k-1}))\},
\end{split}
\end{align}
 to their original model \cite{franke2009validation} introduced in 2009. This additional constraint clearly guarantees the bounds $[0,1]$ of the fractions of chartists and fundamentalists $n^f, n^c$.
Thus, this additional constraint prevents the dynamics from blowing up. We will show in the next section \ref{NumSolv} that this constraint can be rendered redundant by applying an improved time discretization.

\subsection{Robust Formulation of Franke-Westerhoff Model}
\label{NumSolv}
In this section we further analyze the stability problem of the Franke-Westerhoff model in the case additional constraint \eqref{additional} is not enforced.
We first show, that the continuous SDE-ODE system \eqref{ODEFW} is stable without constraint \eqref{additional}.
Based on this result, we show that the stability of the continuous system is preserved for all parameter sets when applying an improved semi-implicit time discretization.

In the previous section, we have shown numerically that the blow up of the  dynamics is caused by the violations of the bounds $[0,1]$ of the agents' fractions.
Proposition 1 shows that these violations are caused by the numerical scheme and are not inherited from the continuous dynamics itself.
 \begin{proposition}\label{propInv}
Any solution of the SDE-ODE system \eqref{ODEFW} remains in the set $V:=\R\times U$ with $U:= \{(x_1,x_2)^T\in\R| x_1\geq 0,\ x_2\geq 0,\  x_1+x_2\leq 1   \}$.
 \end{proposition}
For the proof we refer to appendix \ref{appendixAnalysis}.
Proposition \ref{propInv} shows that the observed blow ups are introduced by the discretization of the SDE-ODE system \eqref{ODEFW}. In order to avoid the instabilities we introduce the following semi-implicit Euler discretization,
\begin{align}
\begin{split}\label{CorrScheme}
&P(t+\Delta t) =P(t)+ \Delta t\ \mu\ D^{FW}(t) + \sqrt{\Delta t}\ \mu  (n^f(t)\ \sigma_f+n^c(t)\ \sigma_c)\ \eta,\quad \eta\sim \mathcal{N}(0,1),\\
& n^f(t+\Delta t) = n^f(t)+ \Delta t\ [  n^c(t+\Delta t) \pi^{cf}(a(t,P,n^f,n^c)) - n^f(t+\Delta t) \pi^{fc}(a(t,P,n^f,n^c))],\\
 &n^c( t+\Delta t) = n^c(t) + \Delta t\ [ n^f(t+\Delta t) \pi^{fc}(a(t,P,n^f,n^c)) -  n^c(t+\Delta t) \pi^{cf}(a(t,P,n^f,n^c))].
\end{split}
\end{align}
 In appendix \ref{appendixModel}, we show that the semi-implicit scheme \eqref{CorrScheme} can be rewritten in explicit form (see equation \eqref{explicitSemiExplicit}). Due to this, the computational cost of the explicit and the semi-implicit schemes are comparable. \textcolor{black}{The scheme satisfies a first order convergence rate. 
This semi-implicit scheme \eqref{CorrScheme} preserves the invariant properties of system \eqref{ODEFW}, which is a strong stability result.}
\begin{proposition}\label{propSol}
For all $\Delta t>0$ and correct initial conditions $(P(t_0), n^f(t_0), n^c(t_0))\in V$ the numerical solution $(P(t_0+k\ \Delta t), n^f(t_0+k\ \Delta t), n^c(t_0+k\ \Delta t))\in V,\ k=0,1,...,N$ defined by the scheme in \eqref{CorrScheme} remains in the set $V$ for any number $N\in\N$ and step size $\Delta t>0$.
\end{proposition}
For the proof we refer to appendix \ref{appendixAnalysis}. This shows that an improved numerical discretization can retain the invariance property of the SDE-ODE system for any time step $\Delta t>0$ and arbitrary choices of constants. This is a huge advantage in comparison of the original model, formulated as a system of difference equations. 
 In particular, the semi-implicit discretization renders the additional constraints \eqref{additional} as introduced by Franke-Westerhoff in \cite{franke2011estimation} redundant.
We conclude with a numerical example of the semi-implicit discretization \eqref{CorrScheme} showing the effectiveness of the semi-implicit time discretization.

\begin{figure}[ht]
    \centering
    \includegraphics[width=\linewidth, height=0.3\linewidth]{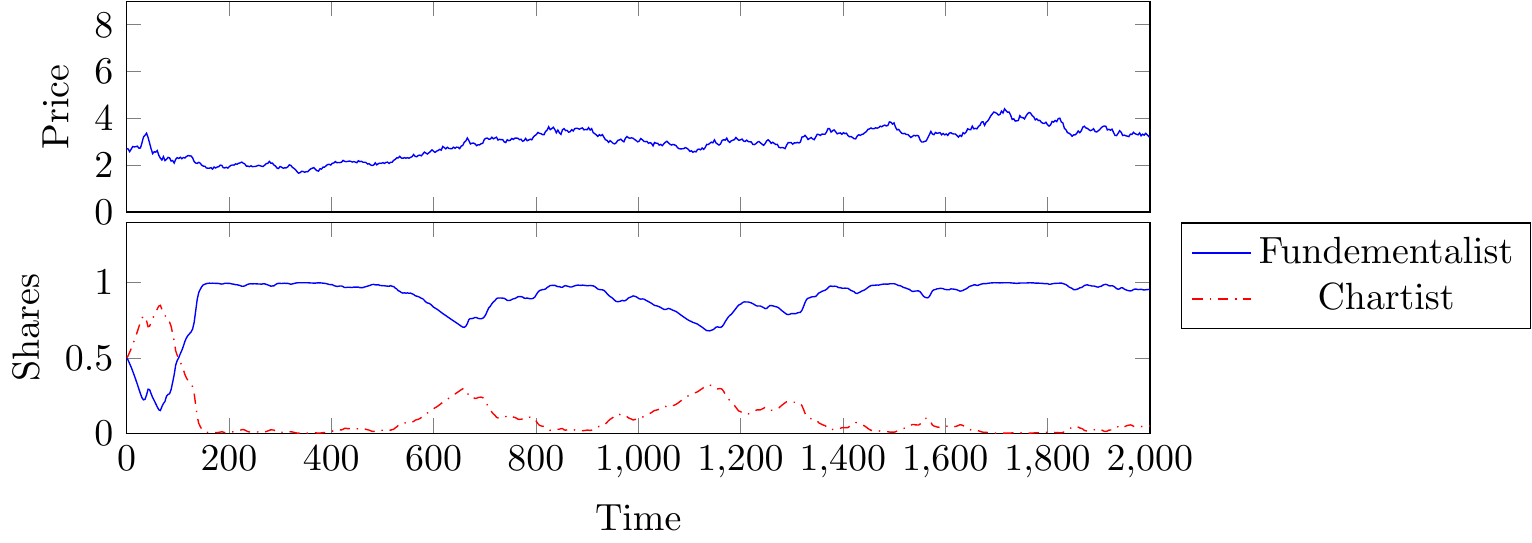}
    \caption{Franke-Westerhoff model with semi-implicit discretization. Parameters as in table \cref{tab:fw_basic} with $\sigma_f=1.15$.}
    \label{fig:tpaci_dt1}
\end{figure}

\begin{figure}[ht]
    \centering
    \includegraphics[width=\linewidth, height=0.3\linewidth]{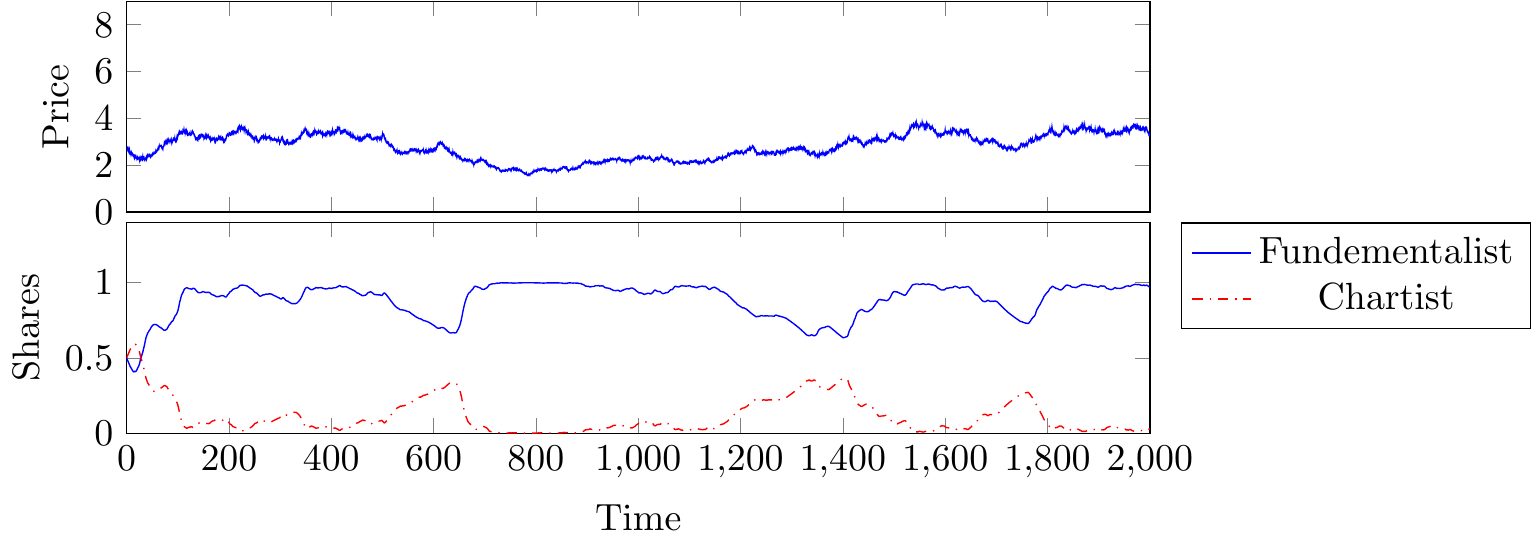}
    \caption{Franke-Westerhoff model with semi-implicit discretization. Parameters as in table \cref{tab:fw_basic} with $\sigma_f=1.15$ and $\Delta t = 0.1$.}
    \label{fig:tpaci_dt0-1}
\end{figure}

\section{Probabilistic Description}\label{PropSection}
\textcolor{black}{
In this section we derive starting from time continuous dynamical system the corresponding mesoscopic description.
More precisely we derive the mean field limit of a simplified version of the Levy-Levy-Solomon model. Furthermore, we 
present a probabilistic description of the Franke-Westerhoff model with the help of the Feynman-Kac formula.
\subsection{Levy-Levy-Solomon Model}
 In this section, we want to derive the mean field limit of a simplified version of the Levy-Levy-Solomon model. We assume that the  investment propensities $\gamma_i$ are constants and the dividend $Z: \R\rightarrow \R_{\geq 0}$ is continuously differentiable and the derivative $ |\dot{Z}|\leq B,\  B>0$ is bounded. Furthermore, we assume that the dividend is not random. 
Thus, the time continuous Levy-Levy-Solomon model for $N$ agents becomes, 
\[
\begin{cases}
& \dot{w}_i(t)=(1-\gamma_i)\ r\ w_i(t)+\gamma_i\ w_i(t)\ \frac{\dot{S}(t)+Z(t)}{S(t)},\\
& w_i(0)=w^0_i,\\
&S(t)=\frac{1}{N} \sum\limits_{k=1}^N \gamma_k\ w_k(t),\quad i=1,...,N.
  \end{cases}
\]
So far we have no explicit ODE system because the right hand side of our differential equation depends on the differential of the stock price. The wealth dynamics can be rewritten in matrix form as,
   \[
 \left(\boldsymbol{I}-\diag(\boldsymbol{\gamma})\frac{\boldsymbol{w}(t)\boldsymbol{\gamma}^T}{\boldsymbol{\gamma}^T\boldsymbol{w}(t)}\right)\ \dot{\boldsymbol{w}}(t)=\diag\big((\boldsymbol{e}-\boldsymbol{\gamma})\ r\big)\ \boldsymbol{w}(t)+ \frac{N}{{\boldsymbol{\gamma}}^T\boldsymbol{w}(t)}\ \diag\big(Z(t)\ \boldsymbol{\gamma}\big)\ \boldsymbol{w}(t),
  \]
where $ \boldsymbol{w}(t)\in\R_{>0}^{N\times 1},\  \boldsymbol{\gamma}=(\gamma_1,\dots,\gamma_N)^T \in (0,1)^N,\ \boldsymbol{e}:=(1,...,1)^T\in \R^N$ and  $\boldsymbol{I}\in\R^{N\times N}$ is the identity matrix. 
We define, 
$$
\boldsymbol{\Sigma}(\boldsymbol{w}(t)):=\left(\boldsymbol{I}-\diag(\boldsymbol{\gamma})\frac{\boldsymbol{w}(t)\boldsymbol{\gamma}^T}{\boldsymbol{\gamma}^T\boldsymbol{w}(t)}\right).
$$ 
In order to derive an  explicit ODE system we need to invert the matrix $\boldsymbol{\Sigma}(\boldsymbol{w}(t))$. Notice that, 
\[
\boldsymbol{\Sigma}(\boldsymbol{w}(t))=\boldsymbol{I}-\boldsymbol{P}(\boldsymbol{w}(t)),
\]
holds with,
\[
\boldsymbol{P}(\boldsymbol{w}(t)):= \begin{pmatrix}  \frac{\gamma_1w_1(t)}{\boldsymbol{\gamma}^T\boldsymbol{w}(t)} \\ \vdots \\ \frac{\gamma_n w_n(t)}{\boldsymbol{\gamma}^T\boldsymbol{w}(t)}
\end{pmatrix}\quad 
\begin{pmatrix}
\gamma_1& \gamma_2 & \dots & \gamma_n
\end{pmatrix}.
\]
The inverse $\boldsymbol{\Sigma}^{-1}(\boldsymbol{w}(t))$ can be computed by a  Neumann series expansion, since $|| \boldsymbol{P}(\boldsymbol{w}(t)) ||_1<1$ holds. 
We get,
$$
\boldsymbol{\Sigma}^{-1}(\boldsymbol{w}(t))=\big(\boldsymbol{I}-\boldsymbol{P}(\boldsymbol{w}(t))\big)^{-1}=\sum\limits_{k=0}^{\infty} \boldsymbol{P}^k(\boldsymbol{w}(t))=\boldsymbol{I}+\frac{1}{1-\alpha}\boldsymbol{P}(\boldsymbol{w}(t)),
$$
with $ \alpha:=\text{trace}\big(\boldsymbol{P}(\boldsymbol{w}(t))\big)=\frac{1}{\boldsymbol{\gamma}^T\boldsymbol{w}(t)}\sum\limits_{i=1}^N \gamma_i^2\ w_i(t).$
Thus, our explicit ODE system reads,
\begin{align}
\begin{split}
 \begin{cases}\label{ExODE}
& \dot{\boldsymbol{w}}(t)=\boldsymbol{\Sigma}^{-1}(\boldsymbol{w}(t))\ \diag\big((\boldsymbol{e}-\boldsymbol{\gamma})\ r\big)\ \boldsymbol{w}(t) + \frac{N}{{\boldsymbol{\gamma}}^T\boldsymbol{w}(t)}\ \boldsymbol{\Sigma}^{-1}(\boldsymbol{w}(t))\ \diag\big(Z(t)\ \boldsymbol{\gamma}\big)\ \boldsymbol{w}(t),\\
& \boldsymbol{w}(0)=\boldsymbol{w}^0 ,
  \end{cases}
  \end{split}
\end{align}
 and for the i-th agent we get, 
\[
\dot{w}_i(t)={ r w_i(t)+\frac{\gamma_iw_i(t)\ Z(t)}{\frac{1}{N} \sum\limits_{j=1}^N (1-\gamma_j)\gamma_jw_j(t)  }},\quad w_i(0)=w_i^0.
\]
\begin{proposition}\label{PropExUn}
 We assume that the dividend $Z: \R\rightarrow \R_{\geq 0}$ is continuously differentiable and the derivative $ |\dot{Z}|\leq B,\  B>0$ is bounded. Then the ODE system \eqref{ExODE} has a unique solution on $[0,\infty|)$. 
\end{proposition}
For a detailed proof we refer to the Appendix \ref{appendixAnalysis}. 
\subsubsection{ Mean Field Limit of Simplified Levy-Levy-Solomon Model}
We employ the empirical measure to derive the  mean field limit equation of our ODE system \eqref{ExODE}. The idea is to connect the solution of the system \eqref{ExODE} to solutions of the mean field equation. 
The empirical measure associated to the solutions\\
 $X_N:=\big(w_1(t),\gamma_1,w_2(t),\gamma_2,\dots,w_N(t) ,\gamma_N \big),\ t\in [0,\infty)$ is,
\[
\mu^N_{X_N}: [0,\infty)\rightarrow \mathcal{P}( \R_{>0}\times(0,1)),\quad \mu^N_{X_N}(t):=\mu^N_{X_N}(t,w,\gamma )=\frac{1}{N}\sum\limits_{k=1}^N \delta (w-w_k(t))\ \delta(\gamma-\gamma_k),
\]
 where $\mathcal{P}( \R_{>0}\times (0,1))$ denotes the space of probability measures on $\R_{>0}\times (0,1)$. 
We consider a test function $\phi\in C_c^1(\R_{>0}\times (0,1) )$ and compute, 
\begin{align*}
\frac{d}{dt}\langle \mu^N_{X_N}(t),\phi(w,\gamma)\rangle &=\frac1N \sum\limits_{k=1}^N \frac{d}{dt} \phi(w_k(t),\gamma_k)=\frac1N \sum\limits_{k=1}^N \p{}{w}\phi(w_k(t),\gamma_k)\cdot \dot{w_k}(t)\\
&= \frac1N \sum\limits_{k=1}^N \p{}{w}\phi(w_k(t),\gamma_k)\cdot\Bigg(  r\ w_k(t)+\ \frac{\gamma_k\ w_k(t) Z(t)}{\frac{1}{N} \sum\limits_{j=1}^N (1-\gamma_j)\ \gamma_j\ w_j(t)}\Bigg)\\
&=\left\langle \mu^N_{X_N}(t), \p{}{w}\phi(w,\gamma)\cdot r\ w\right\rangle\\
&\quad +\left\langle \mu^N_{X_N}(t),\ \p{}{w}\phi(w,\gamma)\cdot \frac {\gamma\ w \ Z(t)}{\frac{1}{N} \sum\limits_{j=1}^N (1-\gamma_j)\ \gamma_j\ w_j(t)}\right\rangle.
\end{align*}
Here, we have used $\langle\cdot\rangle$ as shorthand notation for the $L_2$ inner product on  $\R_{>0}\times (0,1)$. We define,
\begin{align*}
F[\mu^N_{X_N}](t):&=\frac{1}{N} \sum\limits_{k=1}^N (1-\gamma_k)\ \gamma_k\ w_k(t)\\
&=\langle \mu^N_{X_N}(t), (1-\gamma)\ \gamma\ w \rangle\\
&= \int\limits_{\R_{>0}}\int\limits_{0}^1 w\ \gamma\ (1-\gamma)\  \mu^N_{X_N}(\gamma,w,t)\ d\gamma dw, 
\end{align*}
and we thus get,
\begin{align*}
\frac{d}{dt}\langle \mu^N_{X_N}(t),\phi(\gamma, w)\rangle = \left\langle \mu^N_{X_N}(t), \p{}{w}\phi(w,\gamma)\ \left[  r\ w+  \frac{\gamma\ w\ Z(t)}{F[\mu^N_{X_N}](t)}\right]\right\rangle.
\end{align*}
Then, we integrate by parts and obtain,
\begin{align}\label{weakMF}
\left\langle \p{\mu^N_{X_N}}{t} + \p{}{w}\left( \left[  r\ w+  \frac{\gamma\ w\ Z(t)}{F[\mu^N_{X_N}](t)}\right]\  \mu^N_{X_N}\right) , \phi \right\rangle=0.
\end{align}
Equation \eqref{weakMF} is the weak form of the mean field PDE,
\begin{align}\label{MFstrong}
\p{}{t} f(t,w,\gamma)+\p{}{w}\left( r\ w\ f(t,w,\gamma)+\frac{\gamma\ w\  Z(t)}{\int  w^{\prime}\ \gamma^{\prime}\ (1-\gamma^{\prime})\ f(t,w,\gamma)\ dw^{\prime}d\gamma^{\prime} }  f(t,w,\gamma)\right)=0. 
\end{align}
For a rigorous prove of the mean field limit, we apply the method originally used by Dobrushin \citep{dobrushin1979vlasov}.
The key idea is to derive a stability estimate which enables to prove the convergence of the ODE system \eqref{ExODE} to the mean field equation \eqref{MFstrong}. 
We follow the presentation by Golse \cite{golse2016dynamics}. The convergence is obtained in the space of probability measures $\mathcal{P}$ and we use the Wasserstein distance.
For our purposes we solely need the 1-Wasserstein distance which is defined as follows
\begin{definition}
Let $\mu$ and $\nu$ two probability measures on $\R_{>0}\times (0,1)$. Then, the 1-Wasserstein distance $\dist_{W}(\nu,\mu)$ between the measures $\nu$ and $\mu$ is defined by,
\[
\dist_{W}(\nu,\mu):=\inf\limits_{\pi\in\Pi(\nu,\mu)} \int\limits_{\R\times\R} |x-y|\ \pi(dxdy),
\] 
where $\Pi(\nu,\mu)$ is the space of probability measures on $\R_{>0}\times (0,1)\times \R_{>0}\times (0,1)$ such that the marginals of $\pi$ are $\mu$ and $\nu$, respectively, i.e.
$\int d\pi(\cdot,y)=d\mu(\cdot)$ and $\int d\pi(x,\cdot)= d\nu(\cdot)$.
\end{definition}
Furthermore, we introduce the push-forward notion for a measurable map $\Phi: \R_{>0}\times (0,1)\to \R_{>0}\times (0,1)$ and measure $\mu\in\mathcal{P}(\R_{>0}\times (0,1))$.
The measure $\nu\in \mathcal{P}(\R_{>0}\times (0,1)) $ is denoted by $\nu=\Phi\#\mu$, if
$$
\nu(B)=\mu(\Phi^{-1}(B)),
$$
holds for any set $B\subset \R_{>0}\times (0,1)$. Notice that the mean field equation \eqref{MFstrong} can be rewritten for probability measures and that the solutions need to be understood in the sense of distributions. We are ready to proof the mean field limit of the ODE system \eqref{ExODE} to the mean field equation \eqref{MFstrong}.
\begin{theorem}
 Let $f^0$ be a probability density on $\R_{>0}\times (0,1)$ such that, 
\[
0<\int\limits_{\R\times (0,1)} w\ \gamma\  (1-\gamma)\ f^0(w,\gamma)\ dwd\gamma < \infty.
\]
Then the Cauchy problem for the mean field PDE,
\begin{align*}
\begin{cases}
&\p{}{t} f(t,w,\gamma)+\p{}{w}\left( r\ w\ f(t,w,\gamma)+\frac{\gamma\ w\  Z(t)}{\int  w^{\prime}\ \gamma^{\prime}\ (1-\gamma^{\prime})\ f(t,w,\gamma)\ dw^{\prime}d\gamma^{\prime} }  f(t,w,\gamma)\right)=0,\\
&f(0,w,\gamma)=f^0,
\end{cases}
\end{align*}
 $w\in\R_{>0}, \gamma\in (0,1),\ t\in[0,T],\ T\in \R_{>0}$, has a unique weak solution in $C(\R_{>0}\times (0,1),L^1(\R_{>0}\times (0,1)))$. Then the ODE system with initial conditions $X^0_N=(w_1^0,\gamma_1,,...,w_N^0,\gamma_N)$   
\begin{align*}
\begin{cases}
& \dot{w}_i(t)={ r w_i(t)+\frac{\gamma_iw_i(t)\ Z(t)}{\frac{1}{N} \sum\limits_{j=1}^N (1-\gamma_j)\gamma_jw_j(t)  }},\\
& w_i(0)=w_i^0,\\
\end{cases}
\end{align*} 
where $\gamma_i\in (0,1),\ w_i\in \R_{>0}$ has a unique solution denoted by $ X_N=(w_1(t),\gamma_1,...,w_N(t),\gamma_N),\  t\in [0,T]$. If the empirical measure,
\[
\mu_{X_N^0}=\frac{1}{N} \sum\limits_{i=1}^N \delta(w-w_i(0))\ \delta(\gamma-\gamma_i),
\]
satisfies,
\[
\dist_{W}(\mu_{X_N^0},f^0)\rightarrow 0,\quad \text{as}\ N\rightarrow \infty,
\]
then,
\[
\mu_{X_N}\rightharpoonup f(t,\cdot)\mathcal{L},\quad \text{as}\ N\rightarrow \infty,
\]
in the weak topology of probability measures, where $\mathcal{L}$ denotes the Lebesgue measure on $\R_{>0}\times (0,1)$. Furthermore, the convergence rate is given by,
\[
\dist_{W}(\mu_{T_tX_N^0},f(t,\cdot)\mathcal{L})\leq e^{M\ t } \dist_{W}(\mu_{X_N^0},f^0)\rightarrow 0,
\]  
as $N\rightarrow\infty$ for each $t\in [0,T]$.
\end{theorem}
We aim to show the advantages of such a mesoscopic description in comparison to the microscopic agent dynamics.
As short example we present the possibility to derive an explicit solution of the mean field equation \ref{MFstrong}.
\paragraph{Explicit Solution}
An explicit solution can be derived with the following ansatz, 
$$
f(t,w,\gamma) =  \frac{c}{w}\exp\{ - (\ln(w)-B(t,\gamma))^2\} .
$$
Here $c>0$ denotes a normalization constant and the function $B(t,\gamma)$ can be easily obtained by plugging the ansatz in equation \eqref{MFstrong}. 
Thus, $B$ is given by,
$$
B(t,\gamma,A) = \int\limits_0^t\left( r+ \frac{\gamma\ Z(s)}{A(s)}\right)\ ds + \tilde{c},\quad \tilde{c}\in\R. 
$$
Notice that $B$ implicitly depends on the function $f$ by the operator,
$$
A(t):=\int\int w'\ \gamma' (1-\gamma' )\ f(t,\gamma', w')\  dw'd\gamma'. 
$$
Hence, in order to obtain an explicit solution the fixed point equation,
$$
 A(t):=\int\int w'\ \gamma' (1-\gamma' )\ \frac{c}{w}\exp\{ - (\ln(w)-B(t,\gamma,A))^2\}\  dw'd\gamma',
$$
needs to be solved for each time $t$. Thus, we can deduce that the solution is of the type of a log-normal probability distribution. }
%\begin{align*}
%&\frac{d}{dt} f(t,w,\gamma) =  \frac{c}{w}\exp\{ - (\ln(w)-B(t,\gamma))^2\}  -2(\ln(w)-B(t,\gamma))-\frac{d}{dt} B(t,\gamma)\\
%&\partial_w f(t,w,\gamma) = -\frac{1}{w}f(t,w,\gamma)+ \frac{c}{w}\exp\{ - (\ln(w)-B(t,\gamma))^2\}  -2(\ln(w)-B(t,\gamma)) \frac{1}{w}\\
%&\partial_w ([r+\frac{\gamma D(t)}{A}]w    f(t,w,\gamma)) = [r+\frac{\gamma D(t)}{A}] f(t,w,\gamma) + [r+\frac{\gamma D(t)}{A}] w\ \partial_w(f(t,w,\gamma))
%\end{align*}

\subsection{Franke-Westerhoff Model}

\textcolor{black}{
We consider the following time continuous SDE-ODE model,
\begin{align}
\begin{split}
&dP =\textcolor{black}{\mu \frac{ n^f\phi\ (F-P)}{1-\mu\  n^c\ \chi}\ dt+\ \mu  \frac{n^f\ \sigma_f+n^c\ \sigma_c}{1-\mu \ n^c\ \chi}}\ dW,\\
& \frac{d}{dt} n^f = n^c \pi^{cf}(a) - n^f \pi^{fc}(a),\\
 &  \frac{d}{dt} n^c= n^f \pi^{fc}(a) -  n^c \pi^{cf}(a).
\end{split}
\end{align}
Notice that we have reformulated the naive time continuous limit into an explicit ODE-SDE system. In order to ensure well-posedness we assume that $\chi\in(0,1)$ holds.
As discussed for the Levy-Levy-Solomon model in section \ref{CLimit} we emphasize that our choice of time scaling is one particular of several possible.
For simplicity we reduce the dimensions of the system by defining
$$
n= n^f-n^c. 
$$
Then the dynamics can be immediately rewritten as,
\begin{align}\label{expFW}
\begin{split}
&dP =\mu \frac{ \frac{1+n}{2}\phi\ (F-P)}{1- \mu\ \frac{1-n}{2}\ \chi}\ dt+\ \mu  \frac{\frac{1+n}{2}\ \sigma_f+\frac{1-n}{2}\ \sigma_c}{1-\mu\ \frac{1-n}{2} \chi}\ dW,\\
& \frac{d}{dt} n =  (1-n) \pi^{cf}(a(P,n)) -(n+1) \pi^{fc}(a(P,n)).
\end{split}
\end{align}
The the right hand side of the SDE-ODE system\eqref{expFW} is continuously differentiable and thus locally Lipschitz continuous. 
This immediately gives the existence of a unique solution on a finite time interval. The following Proposition is a simple application of the famous Feynman-Kac formula \cite{lapeyre2003introduction}.
\begin{proposition}
We assume that the stochastic process defined by the differential system \eqref{expFW} has a probability distribution function which is a $C^2$ function. We denote the corresponding probability distribution function
by $f(t,p,n),\ p,n\in\R$. Then $f$ satisfies the following Fokker-Planck type equation,
\begin{align}
\begin{split}\label{FWPDE}
&\partial_t f(t,p,n) + \partial_p \Big( \Big[\mu \frac{ \frac{1+n}{2}\phi\ (F-p)}{1- \mu\ \frac{1-n}{2}\ \chi} \Big] f(t,p,n)\Big) \\
&+ \partial_n \Big(\Big[ (1-n) \pi^{cf}(a(p,n)) -(n+1) \pi^{fc}(a(p,n)) \Big] f(t,p,n) \Big)= \frac{\mu^2 }{2} \frac{(\frac{1+n}{2}\ \sigma_f+\frac{1-n}{2}\ \sigma_c)^2}{(1-\mu\ \frac{1-n}{2} \chi)^2}\partial_{pp}^2 f(t,p,n).
\end{split}
\end{align}
\end{proposition}
The PDE \eqref{FWPDE} can be analyzed with respect to steady states, convergence rates or phase transitions. Exemplary we study the steady state behavior of an observable quantity.
We consider the weak form of equation \eqref{FWPDE} and test with the function $\psi(n)=\delta(n)$. Thus we obtain the weak form of the function $g(t,p)= f(t,p,n=0)$ which satisfies,
\begin{align}
\begin{split}\label{MargPDE}
&\partial_t g(t,p) + \partial_p \Big( \Big[\mu\ \frac{ \frac{1}{2}\phi\ (F-p)}{1- \frac{\mu}{2}\ \chi} \Big] g(t,p)\Big) = \frac{\mu^2 }{8} \frac{(\sigma_f+\sigma_c)^2}{(1-\frac{\mu}{2} \chi)^2}\partial_{pp}^2 g(t,p).
\end{split}
\end{align}
For simplicity, we define the constants,
$$
\widehat{\phi}:= \frac{ \frac{\mu}{2}\phi}{1- \frac{\mu}{2}\ \chi},\quad \sigma^2:= \frac{\mu^2}{4}\frac{(\sigma_f+\sigma_c)^2}{(1-\frac{\mu}{2} \chi)^2},
$$
and thus the corresponding steady state equation reads,
$$
 \partial_p \Big( \Big[ \widehat{\phi}\ (F-p) \Big] g_{\infty}(p)\Big) = \frac{\sigma^2 }{2} \partial_{pp}^2 g_{\infty}(p).
$$
The steady state solution is then of Gaussian type and given by,
$$
g_{\infty}=\widehat{c} \exp\left\{ -\frac{\widehat{\phi}}{\sigma^2} (p^2-p\ F) \right\},
$$
with normalization constant $\widehat{c}>0$.
}

\section{Conclusion}
\label{sec-conclusion}

In this study, we derived the time-continuous formulations of the difference equations used in the Levy-Levy-Solomon and the Franke-Westerhoff model. On the example of the Levy-Levy-Solomon model, we showed that these continuous formulations are not unique. Then, we showed that the numerical instabilities present in the standard Franke-Westerhoff model are not present in the time-continuous SDE-ODE system but stem from the explicit Euler discretization and can be alleviated by applying a proper semi-implicit Euler discretization. 
In this manuscript, we showed the immanent importance of the proper choice of numerical discretization to the model behavior. As a consequence, we strongly recommend to model ABCEM models on the continuous level as this allows for studying different time scales and the effects of different time discretizations, some of which may be suitable to overcome additional constraints for stability. Furthermore, a continuous formulation allows for the derivation of PDE models which may be simpler to analyze than the original ABCEM models. \textcolor{black}{Furthermore, we have given an introduction to kinetic theory and have presented the mean field limit. We have derived the mesoscopic description of variants of the Levy-Levy-Solomon and Franke-Westerhoff model. Especially, we were able to characterize the shape of the solution of the mesoscopic Levy-Levy-Solomon model  and the steady state distribution of the Franke-Westerhoff model. These examples already reveal the opportunities to derive analytical results of mesoscopic models. In addition, we can expect that a mesoscopic description usually reduces the costs in comparison to the microscopic system remarkably.  }

\section*{Acknowledgement}
Funded by the Deutsche Forschungsgemeinschaft (DFG, German Research Foundation) under Germany's Excellence Strategy – EXC-2023 Internet of Production – 390621612. This research was support by the RWTH Start-Up initiative. T. Trimborn gratefully acknowledges the support by the Hans-Böckler-Stiftung.

\appendix

\section{Appendix}
\label{sec-appendix}

\subsection{Analysis}
\label{appendixAnalysis}

The proof of the Proposition \ref{propInv} is given by:
\begin{proof}
Since we consider a stochastic It\^o integral it is clear that $P$ remains in $\R$. It remains to show that $U:= \{(x_1,x_2)^T\in \R| x_1\geq 0,\ x_2\geq 0,\  x_1+x_2\leq 1   \}$ is an invariant set  for the equations \eqref{agentdynC}. We define, 
\begin{align*}
&f_1(x_1,x_2) : = x_2\pi^{cf} - x_1 \pi^{fc},\\
 &f_2(x_1,x_2):= x_1 \pi^{fc} -  x_2 \pi^{cf},
\end{align*}
 and show the invariance of $U^+:= \{(x_1,x_2)^T\in\R| x_1\geq 0,\ x_2\geq 0  \}$.
We directly obtain that $f_1(0,x_2)=x_2\pi^{cf}>0$ and $f_2(x_1,0)=x_1\pi^{fc}>0$ holds and  thus $U^+$ is positive invariant. 
Secondly, we define $\phi(x_1,x_2)=x_1+x_2$ and compute the Lie derivative of $f$ along $\phi$,
$$
\langle  \nabla \phi(x_1,x_2) , f(x_1,x_2)\rangle= x_2\pi^{cf} - x_1 \pi^{fc}+ x_1 \pi^{fc} -  x_2 \pi^{cf}=0.
$$
Hence, we have shown the positive invariance  of the set $\{(x_1,x_2)^T \in \R | \phi(x)\leq 1 \}$ and consequently $U$ is an invariant set of our ODE system \eqref{agentdynC}. 

\end{proof}

The proof of the Proposition \ref{propSol} reads:
\begin{proof}
Since the Euler-Maruyama method updating rule is simply a sum of real numbers and the real line is a closed set for any stock price, $P\in\R$ holds. 
The updating rule of the agents' fraction can be rewritten as follows,
\begin{align*}
n^f(t_0+(k+1)\Delta t) &= \frac{n^f(t_0+k\Delta t)+\Delta t \pi^{cf}(a(t_0+k \Delta t)) }{1+\Delta t (\pi^{fc}(a(t_0+k\ \Delta t)) + \pi^{cf}(a(t_0+k\ \Delta t)))},\\
n^c(t_0+(k+1)\Delta t) &= \frac{n^c(t_0+k \Delta t)+\Delta t \pi^{fc}(a(t_0+k\ \Delta t)) }{1+\Delta t (\pi^{fc}(a(t_0+k\ \Delta t)) + \pi^{cf}(a(t_0+k\ \Delta t)))}. 
\end{align*}
Then, we can perform a simple induction. We assume that $0< n^f(t_0+k\Delta t)<1,\ 0< n^c(t_0+k\Delta t) <1$ holds and obtain,
\begin{align*}
n^f(t_0+(k+1)\Delta t) &= \frac{n^f(t_0+k\Delta t)+\Delta t \pi^{cf}(a(t_0+k \Delta t)) }{1+\Delta t (\pi^{fc}(a(t_0+\Delta t)) + \pi^{cf}(a(t_0+k\ \Delta t)))}\\
&\leq \frac{n^f(t_0+k\Delta t)+\Delta t \pi^{cf}(a(t_0+k \Delta t))+ n^c(t_0+k \Delta t)+\Delta t \pi^{fc}(a(t_0+k\ \Delta t))}{1+\Delta t (\pi^{fc}(a(t_0+k\ \Delta t)) + \pi^{cf}(a(t_0+k\ \Delta t)))} \\
&=1,\\
n^c(t_0+(k+1)\Delta t) &= \frac{n^c(t_0+k \Delta t)+\Delta t \pi^{fc}(a(t_0+k\ \Delta t)) }{1+\Delta t (\pi^{fc}(a(t_0+k\ \Delta t)) + \pi^{cf}(a(t_0+k\ \Delta t)))}\\
&\leq  \frac{n^f(t_0+k\Delta t)+\Delta t \pi^{cf}(a(t_0+k \Delta t))+n^c(t_0+k \Delta t)+\Delta t \pi^{fc}(a(t_0+k\ \Delta t)) }{1+\Delta t (\pi^{fc}(a(t_0+k\ \Delta t)) + \pi^{cf}(a(t_0+k\ \Delta t)))}\\
&=1,
\end{align*}
since $\pi^{fc},\pi^{cf}>0$ holds by definition and we have used that  $n^f+n^c=1$ holds. The previous inequality shows that $n^f(t_0+(k+1)\Delta t),n^c(t_0+(k+1)\Delta t)$ remain in the set $U$
for all $k\in\N$. 
\end{proof}

The proof of Proposition \ref{PropExUn}. 
\textcolor{black}{
\begin{proof}
The potential singularity in the second term of the right-hand side of equation \eqref{ExODE} might cause difficulties. We need to ensure that,
\[
\frac{1}{N} \sum\limits_{j=1}^N (1-\gamma_j)\gamma_jw_j(t) >0,\quad \forall\ t\in[ 0,\infty),
\]
holds to obtain existence and uniqueness on $t\in[0,\infty)$.
We define for an arbitrary but fixed number of agents $N\in\N$ the function,
\[
F:\ \R_{>0}^N\times [0,\infty)\rightarrow\R^N,\ (w_1,...,w_N,t)^T \mapsto\begin{pmatrix} r w_1+\frac{\gamma_1w_1\ Z(t)}{\frac{1}{N} \sum\limits_{j=1}^N (1-\gamma_j)\gamma_jw_j  }\\ \vdots\\r w_N+\frac{\gamma_Nw_N\ Z(t)}{\frac{1}{N} \sum\limits_{j=1}^N (1-\gamma_j)\gamma_jw_j }  \end{pmatrix}. 
\]
The function $F(\boldsymbol{w},t)$ is continuously differentiable for all $\boldsymbol{w}=(w_1,...,w_N)^T$. Thus, the function $F(\boldsymbol{w},t)$ is locally Lipschitz. We have for $i,j=1,...,N$,
\begin{align*}
DF_{i,j}=-\frac{\gamma_i w_i Z(t) \frac1N \gamma_j(1-\gamma_j)}{\left( \frac1N \sum\limits_{k=1}^N (1-\gamma_k)\gamma_k w_k  \right)^2}+\delta_{{i j}} \left( r+\frac{\gamma_j Z(t)}{\frac1N \sum\limits_{k=1}^N (1-\gamma_k)\gamma_k w_k}\right).
\end{align*}
We define a weighted $||\cdot||_{1w}$ norm,
\[
||\boldsymbol{x}||_{1w}:= \frac{1}{N}\sum\limits_{i=1}^N \gamma_i\ (1-\gamma_1)\ |x_i|,\ \boldsymbol{x}\in\R^{N\times 1}.
\]
Thus, we can rewrite the Matrix $\boldsymbol{DF}$,
\[
DF_{i,j}=-\frac{\gamma_i w_i Z(t) \frac1N \gamma_j(1-\gamma_j)}{|| \boldsymbol{w}||_{1w}^2}+\delta_{{i j}} \left( r+\frac{\gamma_j Z(t)}{|| \boldsymbol{w}||_{1w}}\right).
\]
We get,
\begin{align*}
||\boldsymbol{DF}||_1=\max\limits_{||\boldsymbol{x}||_{1w}=1} ||\boldsymbol{DF}\ \boldsymbol{x} ||_{1w}=  \max\limits_{j=1,...,N}\frac1N \sum\limits_{i=1}^N \gamma_i\ (1-\gamma_i)\ |DF_{i,j}|.
\end{align*}
Then, we can immediately deduce the inequality,
\[
||\boldsymbol{DF}||_1\leq Z(t)\ \left( \frac{|| \boldsymbol{w}||_{1w}}{|| \boldsymbol{w}||^2_{1w}}+\frac{1}{|| \boldsymbol{w}||_{1w}}  \right)+r,
\]
and obtain as Lipschitz constant,
\[
L:= \left( r\ + \max\limits_{t \in [0,t_1]} Z(t)\cdot \max\limits_{w\in (a,a+\epsilon)^N} \frac{2}{||\boldsymbol{w}||_{1w}}\right).
\]
We have for $\boldsymbol{w},\tilde{\boldsymbol{w}}\in (a,a+\epsilon)^N\subset \R_{>0}^N, \epsilon>0,\  t\in [0,t_1]\subset [0,\infty)$,
\begin{align*}
||F(\boldsymbol{w},t)-F(\tilde{\boldsymbol{w}},t)||\leq\  L\    ||\boldsymbol{w}-\tilde{\boldsymbol{w}}||.
\end{align*}
Obviously, the Lipschitz constant $L$ blows up if we run in the singularity, thus we cannot state a global Lipschitz constant. Furthermore, $L$ is decreasing with respect to $a$ for fixed $\epsilon>0$, which is the lower interval boundary of our wealth variable. The initial conditions of our ODE system are positive $w_i^0>0$ and by assumption $\gamma_i\in(0,1)$. Hence, the Picard-Lindel\"of theorem gives us existence and uniqueness on an interval $[0,t_1],\ t_1\in(0,\infty)$. Especially,
\[
\frac{1}{N} \sum\limits_{j=1}^N (1-\gamma_j)\gamma_jw_j(t)\geq \underbrace{\frac{1}{N} \sum\limits_{j=1}^N (1-\gamma_j)\gamma_jw_j(0)}_{=:\Gamma_N}>0,\quad \forall\ t\in[0,t_1],
\]
holds. Therefore, we can apply the Picard-Lindel\"of theorem iteratively on the intervals $[t_k,t_{k+1}], k\geq 1$. The Lipschitz constant is not increasing for each interval (growth of dividend is bounded). Hence, we get inductively existence and uniqueness on $[0,\infty)$. 
\end{proof}
The proof of Theorem 1.
\begin{proof}
We split the proof in three parts. First we define the mean field characterstic flow and show existence of a unique solution.
This directly gives existence and uniqueness of a solution of our Cauchy problem. Secondly we derive the Dobrushin inequality for our mean field PDE. 
Finally, we connect the mean field PDE with help of the Dobrushin estimate to the microscopic system and proof convergence in the mean field limit. 
%\begin{lemma}\label{flow}
%For each $\xi \in\R_{>0}\times (0,1)$ and each Borel probability measure $\mu^0$, which satisfies
%\[
%0<\int \xi_2^{\prime} (1-\xi_2^{\prime})\ \xi_1^{\prime}\ \mu^0(d\xi_1^{\prime},d\xi_2^{\prime})<\infty,
%\]
%there exist a unique solution denoted by 
%\[
%[0,\infty)\ni t\mapsto W(t,\xi^0,\mu^0)\in\R_{>0}\times (0,1),
%\]
%of class $C^1$ of the problem
%\begin{align*}
%\begin{cases}
%\dot{W}(t,\xi,\mu^0)=\begin{pmatrix} \dot{W}_1(t,\xi,\mu^0)\\ \dot{W}_2(t,\xi,\mu^0) \end{pmatrix}=\begin{pmatrix} r\ W_1(t,\xi,\mu^0)  + \frac{W_1(t,\xi,\mu^0)\ W_2(t,\xi,\mu^0)\ D(t)}{\int \xi_2^{\prime}  (1-\xi_2^{\prime}) \ \xi_1^{\prime} \ \mu(t, d\xi_1^{\prime},d\xi_2^{\prime})}\\ 
%0 \end{pmatrix},\\
%\mu(t)= W(t,\cdot,\mu^0)\# \mu^0,\\
%W(0,\xi,\mu^0)=\xi=\begin{pmatrix} \xi_1\\ \xi_2 \end{pmatrix}.
%\end{cases}
%\end{align*}
%The operator $\#$ is the push-forward of the measure $\mu^0$ under $W$.
%\end{lemma}
\paragraph{Mean Field Characteristic Flow:}
We have,
\[
c:=\int \xi_2^{\prime} (1-\xi_2^{\prime})\ \xi_1^{\prime}\ \mu^0(d\xi_1^{\prime},d\xi_2^{\prime})<\infty,
\]
 and define the weighted $L^1_w$ space,
 \begin{small}
\[
L^1_w:=\left\{p: \R_{>0}\times(0,1)\rightarrow \R^2\ \text{meaurable function},\  \int\limits_{\R_{>0}\times(0,1)} |p(y)|\ y_2 (1-y_2)\ \mu^0(dy_1,dy_2)<\infty \right\}.
\]
\end{small}
The weight is a positive measurable function and thus our space is a Banach space. 
Then, we define the sequence $(W^n)_{n\geq 0}$,
\begin{footnotesize}
\begin{align*}
\begin{cases}
{W}^{n+1}(t,\xi)=\begin{pmatrix} {W}_1^{n+1}(t,\xi)\\ {W}_2^{n+1}(t,\xi) \end{pmatrix}=\begin{pmatrix} \xi_1+ \int\limits_0^t r\ W_1^n(s,\xi)  + \frac{W_1^n(s,\xi)\ W_2^n(s,\xi)\ Z(s)}{\int W_2^n(s,\xi)  (1-W_2^n(s,\xi)) \ W_1^n(s,\xi) \ \mu^0( d\xi_1^{\prime},d\xi_2^{\prime})}\ ds\\ 
\xi_2 \end{pmatrix},\\
W^0(t,\xi)=\xi=\begin{pmatrix} \xi_1\\ \xi_2 \end{pmatrix},\ \xi_2\in(0,1),\ \xi_1>0.
\end{cases}
\end{align*}
\end{footnotesize}
The second component of our ODE system does not change over time, we thus have,
\[
W_2^n(t,\xi)=\xi_2,\ \forall n\in\N.
\]
We use the special structure of our original $N-$particle ODE system to observe,
\begin{align*}
&||W^{n+1}(t,\cdot)-W^n(t,\cdot)||_{L_w^1} =||W_1^{n+1}(t,\cdot)-W_1^n(t,\cdot)||_{L_w^1} \\
&\leq  \int\limits_0^t      r\ ||W_1^{n}(s,\xi)-W_1^{n-1}(s,\xi)||_{L_w^1} \ ds +\\
&\quad \int\limits_0^t \max\limits_{y\in[0,t]} |Z(y)|\ \Big\|   \frac{W_1^n(s,\cdot)}{\int \xi_2^{\prime}  (1-\xi_2^{\prime}) \ W_1^n(s,\xi^\prime) \ \mu^0( d\xi_1^{\prime},d\xi_2^{\prime})}\\
&\quad\quad -\frac{W_1^{n-1}(s,\cdot)}{\int \xi_2^{\prime}  (1-\xi_2^{\prime}) \ W_1^{n-1}(s,\xi^{\prime}) \ \mu^0( d\xi_1^{\prime},d\xi_2^{\prime})} \Big\|_{L_w^1}\ ds\\
&\stackrel{(\star)}{\leq}   \int\limits_0^t \left(r+\max\limits_{y\in[0,t]} |Z(y)|\ L\right)\  ||W_1^{n}(s,\cdot)-W_1^{n-1}(s,\cdot)||_{L_w^1} \ ds\\
&\leq  \int\limits_0^t\dots \int\limits_0^t \left[\left(r+\max\limits_{y\in[0,t]} |Z(y)|\ L\right)\right]^n\ ||W_1^{1}(s,\cdot)-W_1^{0}(s,\cdot)||_{L_w^1} \ ds.
\end{align*}
We have to discuss the local Lipschitz constant $L$. We define the operator,
\[
\hat{F}[g]=\frac{g(\xi)}{\int g(\xi^{\prime}) \xi_2^{\prime} (1-\xi_2^{\prime})\ \mu^0(d\xi_1^{\prime},d\xi_2^{\prime})},
\]
where $g\in L^1_w$. The G$\hat{\text{a}}$taux derivative of this operator is given by,
\begin{align*}
d\hat{F}(g;\psi)=\frac{d}{dh} \hat{F}[g+h\psi]\Big|_{h=0}=&\frac{\psi}{\int g(\xi^{\prime}) \xi_2^{\prime} (1-\xi_2^{\prime})\ 
\mu^0(d\xi_1^{\prime},d\xi_2^{\prime})}-\\
&\frac{g \ \int \psi(\xi^{\prime}) \xi_2^{\prime} (1-\xi_2^{\prime})\ \mu^0(d\xi_1^{\prime},d\xi_2^{\prime})}{\left( \int g(\xi^{\prime}) \xi_2^{\prime} (1-\xi_2^{\prime})\ \mu^0(d\xi_1^{\prime},d\xi_2^{\prime})\right)^2}.
\end{align*}
We have to discuss the operator norm of the  G$\hat{\text{a}}$taux derivative,
\begin{align*}
&\sup\limits_{||\psi||_{L^1_w}=1}\left\| \frac{\psi_1}{\int g(\xi^{\prime}) \xi_2^{\prime} (1-\xi_2^{\prime})\ 
\mu^0(d\xi_1^{\prime},d\xi_2^{\prime})}- \frac{g \ \int \psi(\xi^{\prime}) \xi_2^{\prime} (1-\xi_2^{\prime})\ \mu^0(d\xi_1^{\prime},d\xi_2^{\prime})}{\left( \int g(\xi^{\prime}) \xi_2^{\prime} (1-\xi_2^{\prime})\ \mu^0(d\xi_1^{\prime},d\xi_2^{\prime})\right)^2}\right\|_{L^1_w}\\
&\leq \sup\limits_{||\psi||_{L^1_w}=1} \left\| \frac{\psi}{\int g(\xi^{\prime}) \xi_2^{\prime} (1-\xi_2^{\prime})\ 
\mu^0(d\xi_1^{\prime},d\xi_2^{\prime})}\right\|_{L^1_w}
+\left\| \frac{g \ \int \psi(\xi^{\prime}) \xi_2^{\prime} (1-\xi_2^{\prime})\ \mu^0(d\xi_1^{\prime},d\xi_2^{\prime})}{\left( \int g(\xi^{\prime}) \xi_2^{\prime} (1-\xi_2^{\prime})\ \mu^0(d\xi_1^{\prime},d\xi_2^{\prime})\right)^2}\right\|_{L^1_w}\\
&\leq \sup\limits_{||\psi||_{L^1_w}=1} \frac{ || \psi||_{L^1_w}}{ || g||_{L^1_w}}\ +\frac{||g||_{L^1_w}}{||g||_{L^1_w}^2} \ ||\psi ||_{L^1_w}  
=  \underbrace{\frac{ 2}{ || g||_{L^1_w}} }_{=:L}<\infty.
\end{align*}
for a fixed function $g\in L^1_w$.\klabsatz
If we return to our original inequality $(\star)$, we can apply the previous computations. We observe a local Lipschitz constant $L$. Starting from the initial density $W^0(t,\cdot)\in L^1_w$ which satisfies 
\[
\frac{1}{||W_1^0(t,\cdot)||_{L_w^1}}<\infty,
\]
we can deduce that we never run in the singularity zero. The reason for this is the construction of the sequence, respectively the structure of our ODE. 
The right hand side of our ODE is positive and thus our solution, respectively the sequence, remains positive.\klabsatz 
The dividend is not bounded by definition. Thus,  we can obtain for every time interval $[0,T], \ T\in\R_{>0}$ a constant $M$, 
 \[
 M:=r+\max\limits_{y\in[0,T]} |Z(y)|\ L,\quad  L:= \frac{2}{||W_1^0(t,\cdot)||_{L_w^1}},
 \]
 such that the previous inequality $(\star)$ holds.   
Furthermore, we have the inequality,
\begin{footnotesize}
\begin{align*}
||W^{1}(s,\xi)-W^{0}(s,\xi)||_{L^1_w}=||W_1^{1}(t,\xi)-W_1^{0}(s,\xi)||_{L_w^1}\\
= \int \left| \int\limits_0^s r \xi_1 + \frac{\xi_1   \xi_2\ Z(y)}{\int \xi_2^{\prime} (1-\xi_2^{\prime}) \xi_1^{\prime}\ \mu^0(d\xi_1^{\prime},d\xi_2^{\prime}) } \ dy \right|\ \xi_2 (1-\xi_2)\ \mu^0(d\xi_1,d\xi_2)\\
\leq \int\limits_0^t \int \left| r \xi_1 + \frac{\xi_1  \ Z(y)}{\int \xi_2^{\prime} (1-\xi_2^{\prime})\xi_1^{\prime}\  \mu^0(d\xi_1^{\prime},d\xi_2^{\prime}) } \right|  \xi_2 (1-\xi_2)\  d\mu^0(d\xi_1,d\xi_2) \ dy \\
\leq  \left(rc+\max\limits_{y\in[0,t]} |Z(y)|\ 1\right)\  t.
\end{align*}
\end{footnotesize}
and finally observe,
\[
||W^{n+1}(t,\cdot)-W^n(t,\cdot)||_{L_w^1} \leq \left(rc+\max\limits_{y\in[0,t]} |Z(y)|\right)\ \frac{\left(M \ t \right)^{n}}{n!}.
\]
Hence, we have,
\[
W^n(t,\cdot)\rightarrow W(t,\cdot),
\]
uniformly in $L_w^1$ and $t\in[0,T]$. 
Furthermore, $W\in C(\R_{>0}\times (0,1),L_w^1)$ satisfies,
\begin{align}\label{integral}
{W}(t,\xi)=\begin{pmatrix} {W}_1(t,\xi)\\ {W}_2(t,\xi) \end{pmatrix}=\begin{pmatrix} \xi_1+ \int\limits_0^t r\ W_1(s,\xi)  + \frac{W_1(s,\xi)\ W_2(s,\xi)\ Z(s)}{\int W_2(s,\xi)  (1-W_2(s,\xi)) \ W_1(s,\xi) \ \mu^0( d\xi_1^{\prime},d\xi_2^{\prime})}\ ds\\ 
\xi_2 \end{pmatrix}.
\end{align}
As next step, we prove the uniqueness of our solutions. We assume that $W(t,\xi), \tilde{W}(t,\xi)$ are two solutions. The uniqueness in the second component is obvious because both solutions have to satisfy the initial conditions. In the first component, we deduce the following inequality,
\begin{footnotesize}
\begin{align*}
|| W_1(t,\xi)-\tilde{W}_1(t,\xi)||_{L^1_w}\leq &  \int\limits_0^t r\ ||W_1(s,\xi)-\tilde{W}_1(s,\xi) ||_{L^1_w}\ ds \\
&+\int\limits_0^t  \xi_2 \left\| \frac{Z(s)\ W_1(s,\xi)}{\int W_1(s,\xi^{\prime})\ \xi_2  (1-\xi_2)\ \mu^0(d\xi^{\prime})}-\frac{Z(s)\ \tilde{W}_1(s,\xi)}{\int \tilde{W}_1(s,\xi^{\prime})\ \xi_2  (1-\xi_2)\ \mu^0(d\xi^{\prime})} \right\|_{L^1_w}\ ds\\
\leq & \int\limits_0^t  \left(r+Z(s)\ L\right)\ ||W_1(s,\xi)-\tilde{W}_1(s,\xi)||_{L^1_w}\ ds.
\end{align*}
\end{footnotesize}
Then, we immediately observe uniqueness by applying the Gronwall inequality.\\
The integrant in equation \eqref{integral} is continuous and thus the function $t\mapsto W(t,\xi)$ is of class $C^1$ on $[0,T]$ and satisfies,
\begin{align*}
\begin{cases}
\p{}{t} W(t,\xi) = \begin{pmatrix} r\ W_1(t,\xi)  + \frac{W_1(t,\xi)\ W_2(t,\xi)\ Z(t)}{\int W_2(t,\xi)  (1-W_2(t,\xi)) \ W_1(t,\xi) \ \mu^0( d\xi_1^{\prime},d\xi_2^{\prime})}\\
0 
\end{pmatrix}\\
W(0,\xi)=\xi.
\end{cases}
\end{align*}
We then finally substitute,
\begin{align*}
& w^{\prime} = W_1(t,\xi^{\prime}),\\
& \gamma^{\prime}= W_2(t,\xi^{\prime}),
\end{align*}
and the integral in the denominator is given by,
\begin{align*}
\int W_1(t,\xi^{\prime})\ W_2(t,\xi^{\prime})\ (1-W_2(t,\xi^{\prime}))\ \mu^0(d\xi^{\prime})=\int w^{\prime} \gamma^{\prime} (1-\gamma^{\prime})\ \underbrace{W(t,\cdot)\#\mu^0(d\xi^{\prime})}_{=\mu(t)}.
\end{align*}
Thus, we have shown the existence and uniqueness of a solution of our mean field characteristic equation. Thanks to the method of characteristics we observe,
\[
f(t,\cdot)\mathcal{L}  = W(t,\cdot,f^0\mathcal{L})\#f^0\mathcal{L},\ \forall \ t\in\R.
\]
Thus, we can immediately deduce  uniqueness of the solution of the Cauchy problem in $C(\R_{>0}\times (0,1),L^1(\R))$ for the mean field PDE.\\
\paragraph{Stability Estimate}
%\begin{theorem}
%Let $\mu_1^0$ and $\mu_2^0$ two measures on $\R_{>0}\times (0,1)$ which satisfy
%\begin{align*}
% 0<\int\limits_{\R_{>0}}\int\limits_0^1  \xi_1^{\prime}\ \xi_2^{\prime}\ (1-\xi_2^{\prime}) \       \mu_i^0(d\xi_1^{\prime},d\xi_2^{\prime})<\infty,
%\end{align*}
%$i\in\{1,2\}$. Then, the measures
%\begin{align*}
%\mu_i(t)=W(t,\cdot,\mu_i^0)\#\mu_i^0,
%\end{align*}
%where $W$ is the mean field characteristic flow defined in lemma \ref{flow} satisfy
%\[
%dist_{W,1}(\mu_1(t),\mu_2(t))\leq e^{M\ t}\ dist_{W,1}(\mu_1^0,\mu_2^0),
%\]
%where $M>0$ is a positive constant and $t\in[0,T]$.
%\end{theorem}
Let $\xi,\tilde{\xi}\in\R_{>0}\times (0,1)$, then we have,
\begin{small}
\begin{align*}
& W(t,\xi,\mu_1^0)-W(t,\tilde{\xi},\mu_2^0)=\xi-\tilde{\xi}\\
& \int\limits_0^{{t}} r\ W_1(s,\xi,\mu_1^0)-r\ W_1(s,\tilde{\xi},\mu_1^0)+ \frac{\xi_2 W_1(s,\xi,\mu_1^0) \ Z(s)}{\int \xi_1^{\prime}\ \xi_2^{\prime}\ (1-\xi_2^{\prime}) \   \mu_1(s,d\xi_1^{\prime},d\xi_2^{\prime}) }- \frac{\tilde{\xi}_2 W_1(s,\tilde{\xi},\mu_2^0)\ Z(s) }{\int \tilde{\xi}_1^{\prime}\ \tilde{\xi}_2^{\prime}\ (1-\tilde{\xi}_2^{\prime}) \   \mu_2(s,d\tilde{\xi}_1^{\prime},d\tilde{\xi}_2^{\prime}) }  \ ds .
\end{align*}
\end{small}
Then, we do a substitution in the integrals in the denominator and since $\mu_i(t)=W(t,\cdot,\mu_i^0)\#\mu_i^0$
we observe e.g. for $i=1$,
\begin{align*}
\int \xi_1^{\prime}\ \xi_2^{\prime}\ (1-\xi_2^{\prime}) \   \mu_1(s,d{\xi}_1^{\prime},d{\xi}_2^{\prime})=
\int W_1(s,\xi^{\prime},\mu_1^0)\ \xi_2^{\prime}\ (1-\xi_2^{\prime}) \   \mu_1^0(d{\xi}_1^{\prime},d{\xi}_2^{\prime}).
\end{align*}
We then use the Lipschitz continuity again. By abuse of notation, we use the same notation for our constant although they differ due to different norms we consider,
\begin{align*}
| W(t,\xi,\mu_1^0)-W(t,\tilde{\xi},\mu_2^0)|\leq |\xi-\tilde{\xi}|+ \underbrace{(r+\max\limits_{y\in[0,\tilde{t}]} |Z(y)|\ L)}_{=:M}\ \int\limits_0^{{t}} | W(s,\xi,\mu_1^0)-W(s,\xi,\mu_2^0)|\ ds.
\end{align*}
 Then, we integrate with respect to the measure $\pi^0\in\Pi(\mu_1^0,\mu_2^0) $ and use Fubini,
 \begin{align*}
 \int | W(t,\xi^{\prime},\mu_1^0)-W(t,\tilde{\xi}^{\prime},\mu_2^0)|\ \pi^0(d\xi^{\prime},d\tilde{\xi}^{\prime})\leq  \int | \xi^{\prime}-\tilde{\xi}^{\prime}|\ \pi^0(d\xi^{\prime},d\tilde{\xi}^{\prime})\\
 +M \int\limits_0^{{t}} \int | W(s,\xi^{\prime},\mu_1^0)-W(s,\tilde{\xi}^{\prime},\mu_2^0)|\ \pi^0(d\xi^{\prime},d\tilde{\xi}^{\prime})\ ds.
\end{align*}
We define,
\[
\Xi[\pi^0](t):= \int | W(t,\xi^{\prime},\mu_1^0)-W(t,\tilde{\xi}^{\prime},\mu_2^0)|\ \pi^0(d\xi^{\prime},d\tilde{\xi}^{\prime}).
\]
Thus, we can rewrite the inequality above to be,
\[
\Xi[\pi^0](t)\leq \Xi[\pi^0](0)+M \int\limits_0^{{t}}  \Xi[\pi](s)\ ds,
\]
and can deduce by the Gronwall inequality that,
\[
\Xi[\pi^0](t)\leq \Xi[\pi^0](0)\ e^{M\ {t}},
\]
holds. We then define the map,
\[
\Phi_t: (\xi,\tilde{\xi})\mapsto \big( W(t,\xi,\mu_1^0),W(t,\tilde{\xi}, \mu_2^0) \big),
\]
and 
\begin{align*}
\Phi_t\# \pi^0=\pi(t)\in \Pi(\mu_1(t),\mu_2(t)),
\end{align*}
since $\pi^0\in \Pi(\mu_1^0,\mu_2^0)$ and $W(t,\cdot,\mu_i^0)\#\mu_i^0=\mu_i(t)$ for all $t\in[0,T]$. 
Hence, we have,
\begin{align*}
\text{dist}_{W,1}(\mu_1(t),\mu_2(t))&=\inf\limits_{\pi\in\Pi(\mu_1(t),\mu_2(t))}\int |\xi-\tilde{\xi}|\ \pi(d\xi,d\tilde{\xi})\\
 &	\leq \inf\limits_{\pi^0\in\Pi(\mu_1^0,\mu_2^0)}\int |W(t,\xi,\mu_1^0)-W(t,\tilde{\xi},\mu_2^0)|\ \pi^0(d\xi,d\tilde{\xi})\\
& =\inf\limits_{\pi^0\in\Pi(\mu_1^0,\mu_2^0)} \Xi[\pi^0](t)\\
&\leq e^{M\ t}\ \inf\limits_{\pi^0\in\Pi(\mu_1^0,\mu_2^0)} \Xi[\pi^0](0)\\
& =e^{M\ {t}}\ \text{dist}_{W,1}(\mu_1^0,\mu_2^0), 
\end{align*}
which gives the result. The first inequality is delicate. In essence, one switches from the infimum over the couplings $\Pi(\mu_1(t), \mu_2(t))$ to the infimum 
over $\Pi(\mu_1^0(t), \mu_2^0(t))$ due to the choice of one transformation. Since the set  $\Pi(\mu_1^0(t), \mu_2^0(t))$ is possibly smaller than $\Pi(\mu_1(t), \mu_2(t))$, the infimum becomes larger. \\
\paragraph{Mean Field Limit:}
By Dobrushin's stability estimate we get,
\[
\text{dist}_{W,1}(\mu_{X_N},f(t,\cdot,\cdot)\mathcal{L})\leq e^{M\ t}\ \text{dist}_{W,1}(\mu_{X_N^0},f^0).
\]
By assumption,
\[
\text{dist}_{W,1}(\mu_{X_N^0},f^0)\rightarrow 0,\quad \text{as}\ N\rightarrow \infty,
\]
and thus,
\[
\text{dist}_{W,1}(\mu_{X_N},f(t,\cdot,\cdot)\mathcal{L}) \rightarrow 0,\quad \text{as}\ N\rightarrow \infty,
\]
holds. Hence, 
\[
\mu_{X_N}\rightharpoonup f(t,\cdot)\mathcal{L},\quad \text{as}\ N\rightarrow \infty.
\]
in the weak topology of probability measures.
\end{proof} }

\subsection{Models}
\label{appendixModel}

\paragraph{Levy-Levy-Solomon Model}
We have implemented the model as defined in \cite{levy1994microscopic, levy1995microscopic}. 
As described in section \ref{TimeDiscretization} we have added the correct time scale to the model. In order to obtain the original model one needs to set $\Delta t=1$.\\ \\
The model considers $N\in\N$ financial agents who can invest $\gamma_i\in [0.01, 0.99],\ i=1,...,N$ of their wealth $w_i\in\R_{>0}$ in a stocks and have to invest $1-\gamma_i$ of their wealth in a safe bond with interest rate $r\in(0,1)$. The investment propensities $\gamma_i$ are determined by a utility maximization and the wealth dynamic of each agent at time $t\in[0,\infty )$ is given by,
\begin{footnotesize}
\begin{align*}
&{w}_i(t)=w_i(t-\Delta t)\\
&\quad +\Delta t \left((1-\gamma_i(t-\Delta t))\ r\ w_i(t-\Delta t)+\gamma_i(t-\Delta t)\ w_i(t-\Delta t)\ \underbrace{\frac{\frac{{S}(t)-S(t-\Delta t)}{\Delta t}+Z(t)}{S(t)}}_{=:x(S,t,Z)}\right).
\end{align*} 
 \end{footnotesize}
The dynamics is driven by a multiplicative dividend process, given by,
\[
Z(t):=(1+\Delta t\  \tilde{z})\  Z(t-\Delta t),
\]
where $\tilde{z}$ is a uniformly distributed random variable with support $[z_1,z_2]$.  
The price is fixed by the so called \textit{market clearance condition}, where $n\in\N$ is the fixed number of stocks and $n_i(t)$ the number of stocks of each agent,  
\begin{align}
n=\sum\limits_{i=1}^N n_i(t)=\sum\limits_{k=1}^N \frac{\gamma_k(t)\ w_k(t)}{S(t)}. \label{fixedpointLLS}
\end{align}
The utility maximization is given by,
\[
\max\limits_{\gamma_i \in [0.01,0.99]} E[\log(w(t+\Delta t,\gamma_i,S^h))],
\]
with,
\begin{align*}
E[\log(w(t+\Delta t, \gamma_i,S^h))]=\frac{1}{m_i} \sum\limits_{j=1}^{m_i} U_i\Bigg(&(1-\gamma_{i}(t)) w_{i}(t,S^h) \left(1+r\Delta t\right)\\ 
&+\gamma_{i}(t) w_{i}(t, S^h) \Big(1+x\big(S,t-j\Delta t,Z\big)\ \Delta t\Big)\Bigg).
\end{align*}

The constant $m_i$ denotes the number of time steps each agent looks back. Thus, the number of time steps $m_i$ and the length of the time step $\Delta t$ defines the time period each agent extrapolates the past values. The superscript $h$ indicates, that the stock price is uncertain and needs to be fixed by the market clearance condition.
Finally, the computed optimal investment proportion gets blurred by a noise term,
\[
\gamma_i (t)=H(\gamma_i^{*}(t)+\epsilon_i),
\]
where $\epsilon_i$ is a normally distributed random variable with standard deviation $\sigma_{\gamma}$. The function $H$ ensures that $\gamma_i$ remains in the interval $[0.01, 0.99]$.
Finally, we have to update the price after the noising process. Since the investment fraction is constant we are able to compute the stock price explicitly,
$$
S(t)=\frac{\frac{1}{n} \sum\limits_{i=1}^N\gamma_i(t) \Big(  w_i(t-\Delta t)+\Delta t\  w_i(t-\Delta t)\big(  \gamma_i(t-\Delta t)\frac{Z(t-\Delta t)-S(t-\Delta t)} {  \Delta t\ S(t-\Delta t)   } +(1-\gamma_i(t-\Delta t))\ r  \big) \Big)}{1-\frac{1}{n } \sum\limits_{i=1}^N  \frac{\gamma_i(t)\gamma_i(t-\Delta t) w_i(t-\Delta t)}{S(t-\Delta t)}}.
$$

\paragraph{Utility Maximization}
Thanks to the simple utility function and linear dynamics we can compute the optimal investment proportion in the cases where the maximum is reached at the boundaries.
The first order necessary condition is given by,
$$
f(\gamma_i):= \frac{d}{dt} E[\log(w(t+\Delta t, \gamma_i,S^h))] = \frac{1}{m_i} \sum\limits_{j=1}^{m_i} \frac{\Delta t\ (x\big(S,t-j\Delta t,Z\big)-r)}{\Delta t\ (x\big(S,t-j\Delta t,Z\big)-r)\ \gamma_i+ 1+\Delta t\ r}.
$$
 Thus, for $f(0.01)<0$ we can conclude that $\gamma_i=0.01$ holds. In the same manner, we get $\gamma_i=0.99$, if $f(0.01)>0$ and $f(0.99)>0$ holds. 
 Hence, solutions in the interior of $[0.01, 0.99]$ can be only expected in the case: $f(0.01)>0$ and  $f(0.99)<0$. This coincides with the observations in \cite{samanidou2007agent}.

\paragraph{Franke-Westerhoff Model}
We present the Franke-Westerhoff model as introduced in \cite{franke2009validation} and considered with minor modifications in \cite{franke2011estimation, franke2012structural}. 
As described in section \ref{TimeDiscretization} we have added a time scaling to the model. In order to obtain the original model one needs consider the explicit Euler discretization of the agents' shares and has to set $\Delta t=1$.
The Franke-Westerhoff model considers two types of agents, chartists and fundamentalists. The demand of each agent reads,
\begin{align}
\mathfrak{D}^f (t)= \phi (F-P(t)) + \epsilon_k^f, \quad \phi \in \mathbb{R}^+, \quad \epsilon_k^f \sim \mathcal{N}(0, \sigma_f^2), \\
\mathfrak{D}^c (t)= \chi \frac{P(t) - P(t-\Delta t)}{\Delta t} + \epsilon_k^c, \quad \chi \in \mathbb{R}^+, \quad \epsilon_k^c \sim \mathcal{N}(0, \sigma_c^2),
\end{align}
where $P(t)$ denotes the logarithmic market price and $F$ denotes the fundamental price. The noise terms $\epsilon_k^f$ and $\epsilon_k^c$ are normally distributed, with zero mean and different standard deviations $\sigma_c^2$ and $\sigma_f^2$. The second important features are the fractions of the chartist or fundamental population. In that sense the two agents can be seen as representative agents of a population. The fraction of chartists $n^C(t)\in[0,1]$ and the fraction of fundamentalist $n^F(t)\in[0,1]$ have to fulfill $n^C(t)+n^F(t)=1$. 
Hence, the deterministic excess demand can be defined as,
\begin{equation}
D^{FW}(t) := \frac12 ( d^f(t) + d^c(t)),
\end{equation}
\begin{align}
d^f(t) := 2\ n^f(t) E[\mathfrak{D}^f(t)], \nonumber \\
d^c(t) := 2\ n^c(t) E[\mathfrak{D}^c(t)]. 
\end{align}
Here, $E$ denotes the expected value. The pricing equation is then given by the simple rule,
\begin{equation}
P(t) = P(t-\Delta t) + \mu\ \Delta t\ D^{FW}(t)+ \sqrt{\Delta t}\ \mu\   \textcolor{black}{(n^f(t)\ \sigma_f+n^c(t)\ \sigma_c)}\ \eta,\quad \eta\sim \mathcal{N}(0,1).
\end{equation}
Finally, we need to specify the switching mechanism. This switching mechanism is known as  the transition probability approach (TPA)  \citep{weidlich2012concepts, lux1995herd}.
We consider the so called switching index $a(t,P,n^f,n^c)\in\R$ which describes the attractiveness of the fundamental strategy over the chartist strategy. Thus, a positive $a(t,P,n^f,n^c)$ reflects an advantage of the fundamental strategy in comparison to the chartist and if $a(t,P,n^f,n^c)$ is negative we have the opposite situation. 
We define the switching probabilities,
\begin{align}
\pi^{cf}(a(t,P,n^f,n^c)) &:= \nu  \exp(a(t,P,n^f,n^c)),\\
\pi^{fc}(a(t,P,n^f,n^c)) &:=  \nu  \exp(- a(t,P,n^f,n^c)).
\label{eq:switchProb_fw}
\end{align}
where $\pi^{xy}$ is the probability that an agent with strategy $x$ switches to strategy $y$. The flexibility parameter $\nu>0$ is a scaling factor for $a(t,P,n^f,n^c)$.
\begin{remark}\label{addRemark}
A minor modification of the Franke-Westerhoff model introduced in 2011 \cite{franke2011estimation, franke2012structural} considers the following switching probabilities,
\begin{align}
\label{eq:switchProbWguard_fw}
\pi^{cf}(a(t,P,n^f,n^c)) &:= \min \left(1, \nu  \exp(a(t,P,n^f,n^c))  \right), \\
\pi^{fc}(a(t,P,n^f,n^c)) &:=  \min \left( 1, \nu  \exp(- a(t,P,n^f,n^c)) \right).
\end{align}
The previous definition ensures that switching probabilities are restricted to the interval $[0,1] \subset \mathbb{R}$. 
In the original model introduced in 2009 \cite{franke2009validation} there has been no additional limits as introduced in \eqref{eq:switchProbWguard_fw}.
\end{remark}

\paragraph{Explicit Euler Discretization}
Then the explicit Euler discretization of the time evolution of chartist and fundamentalist shares as presented in \eqref{ODEFW} is given by,
\begin{align*}
n^f(t) &= n^f(t-\Delta t) +\Delta t n^c(t) \pi^{cf}(a(t-\Delta t)) - \Delta tn^f(t) \pi^{fc}(a(t-\Delta t)),\\
n^c(t) &= n^c(t-\Delta t) + \Delta t n^f(t-\Delta t) \pi^{fc}(a(t-\Delta t)) - \Delta t n^c(t-\Delta t) \pi^{cf}(a(t-\Delta t)).
\end{align*}
\paragraph{Semi-Implicit Euler Discretization}
Alternatively one may use the semi-implicit scheme introduced in section \ref{examples} for the time evolution of chartist and fundamentalist shares
\begin{equation}
\label{explicitSemiExplicit}
\begin{aligned}
n^f(t) &= \frac{n^f(t-\Delta t)+\Delta t \pi^{cf}(a(t-\Delta t)) }{1+\Delta t (\pi^{fc}(a(t-\Delta t)) + \pi^{cf}(a(t-\Delta t)))} ,\\
n^c(t) &= \frac{n^c(t-\Delta t)+\Delta t \pi^{fc}(a(t-\Delta t)) }{1+\Delta t (\pi^{fc}(a(t-\Delta t)) + \pi^{cf}(a(t-\Delta t)))}.
\end{aligned}
\end{equation}
As shown in section \ref{TimeDiscretization} as well,  this numerical approximation stable and conserves the invariance property of the ODEs.\\[3em]

Finally, we have to specify how the switching index $a(t,P,n^f,n^c)$ is calculated. The switching index $a(t,P,n^f,n^c)$, encodes how favourable a fundamentalist strategy is over a chartist strategy. 
The switching index is determined linearly out of the the three principles \textit{predisposition, herding and misalignment},
$$
a(t,P,n^f,n^c) =\alpha_p+ \alpha_h\ (n^f(t)-n^c(t))+ \alpha_m\ (P(t)-F)^2,
$$
where $\alpha_p, \alpha_h,\alpha_m>0$ are weights respectively scaling factors. The sign  $\alpha_p\in\R$ determines the predisposition with respect to a fundamental or chartist strategy. 
For details regarding the modeling we refer to \citep{franke2009validation}.

\subsection{Parameter Sets}\label{parameter}

\paragraph{Levy-Levy-Solomon Model}
The initialization of the stock return is performed by creating an artificial history of stock returns. The artificial history is modeled as a Gaussian random variable with mean
$\mu_h$ and standard deviation $\sigma_h$. Furthermore, we have to point out that the increments of the dividend is deterministic, if $z_1=z_2$ holds. We used the C++  standard random number generator for all simulations of the Levy-Levy-Solomon model if not otherwise stated. 
\begin{table}
	\begin{subtable}[b]{0.45\textwidth}
		\begin{center}
			\begin{tabular}{|c||c|}
			\hline
			Parameter & Value\\
			\hline
			\hline
			$N$ & $100$\\
			\hline
			$m_i$& $15$\\
			\hline 
			$\sigma_{\gamma}$ & $ 0 $ or $0.2$\\
			\hline
			$r$& $0.04$ \\
			\hline 
			$z_1=z_2$& $0.05$\\
			\hline
			$\Delta t$ &$ 1$\\
			\hline
			time steps & 200 \\
			\hline
			\end{tabular}
		\end{center}
		\caption{Parameters of Levy-Levy-Solomon model.}
	\end{subtable}
	\hspace{0.5cm}
	\begin{subtable}[b]{0.45\textwidth}
		\begin{center}
			\begin{tabular}{|c||c|}
			\hline
			 Variable & Initial Value\\
			\hline
			\hline
			$\mu_h$ & $0.0415$\\
			\hline
			$\sigma_h$ & $0.003$\\
			\hline
			$\gamma(t=0)$ & $0.4$\\
			\hline
			 $w_i(t=0)$ & $1000$ \\
			 \hline
			 $n_i (t=0)$ & $100$ \\
			 \hline
			 $S (t=0)$ & $4$ \\
			 \hline
			 $Z (t=0)$ & $0.2$ \\
			 \hline
			\end{tabular}
		\end{center}
		\caption{Initial values of Levy-Levy-Solomon model.}
	\end{subtable}
	\caption{Basic setting of the Levy-Levy-Solomon model.} \label{LLS-basic}
\end{table}

\begin{table}
	\begin{subtable}[b]{0.45\textwidth}
		\begin{center}
			\begin{tabular}{|c||c|}
			\hline
			Parameter & Value\\
			\hline
			\hline
			$N$ & $99$\\
			\hline
			$m_i $&  $10,\ 1 \leqslant i \leqslant 33$  \\
			                  & $141,\ 34 \leqslant i \leqslant 66$ \\
			                  & $256,\ 67 \leqslant i \leqslant 99$ \\
			\hline 
			$\sigma_{\gamma}$ & $ 0.2 $\\
			\hline
			$r$& $0.0001$ \\
			\hline 
			$z_1=z_2$& $0.00015$\\
			\hline
			$\Delta t$ &$ 1 $\\
			\hline 
			time steps & $20,000$\\
			\hline
			\end{tabular}
		\end{center}
		\caption{Parameters of Levy-Levy-Solomon model.} 
	\end{subtable}
	\hfill
	\begin{subtable}[b]{0.45\textwidth}
		\begin{center}
			\begin{tabular}{|c||c|}
			\hline
			 Variable & Initial Value\\
			\hline
			\hline
			  $\mu_h$ & $0.0415$ \\
			  \hline
			  $\sigma_h$ &$ 0.003$ \\
			  \hline
			  $\gamma_i(t=0)$ & $0.4$ \\
			  \hline
			  $w_i (t=0)$ & $1000$ \\
			  \hline
			   $n_i (t=0)$ & $100$ \\
			   \hline
			    $S (t=0)$ & $4 $\\
			    \hline
			   $Z (t=0)$ & $0.004$ \\
			 \hline
			\end{tabular}
		\end{center}
		\caption{Initial values of Levy-Levy-Solomon model.}
	\end{subtable}
	\caption{Setting for the Levy-Levy-Solomon model (3 agent groups).}
	\label{LLS-3-agents}
\end{table}

\paragraph{Franke-Westerhoff Model}

\begin{table}[h!]
    \begin{subtable}[b]{0.45\textwidth}
        \begin{center}
            \begin{tabular}{|c||c|}
                \hline
                Parameter & Value\\
                \hline
                \hline
                $\phi$ & $0.18$\\
                \hline
                $\chi$ & $2.3$\\
                \hline
                $\alpha_0$ & $-0.161$ \\
                \hline
                $\alpha_h$ & $1.3$ \\
                \hline 
                $\alpha_m$ & $12.5$\\
                \hline
                $\sigma_f$ & $0.79$\\
                \hline  
                $\sigma_c$ & $1.9$\\
                \hline
                $\nu$ & $0.05$ \\
                \hline
                $P_f$ & $1$ \\
                \hline
                $\mu$ & $0.01$\\
                \hline
                $\Delta t$ & $1$\\	            
                \hline
            \end{tabular}
        \end{center}
        \caption{Parameters}
    \end{subtable}
    \hfill
    \begin{subtable}[b]{0.45\textwidth}
        \begin{center}
            \begin{tabular}{|c||c|}
                \hline
                Variable & Initial Value\\
                \hline
                \hline
                $P(t=0)$ & 1 \\
                \hline	            
            \end{tabular}
        \end{center}
        \caption{Initial Values}
    \end{subtable}
    \caption{Parameters and initial values  for the Franke-Westerhoff model.}
    \label{tab:fw_basic}
\end{table}

\begin{table}
    \centering
    \begin{tabular}{|c||c|}
        \hline
        Simulation & Random Number Generator \\
        \hline          
        \hline
                \Cref{LLS-DeltaT} & C++ MT19937 RNG ($64$ bit)\\	            
        \hline
                \Cref{LLS-DeltaTfixedM} & C++ MT19937 RNG ($64$ bit)\\	            
        \hline
                \Cref{fig:tpac_dt1_noblowup} & IntelMKL MT2203 RNG ($64$ bit)\\	            
        \hline
                \Cref{fig:tpac_dt1} & IntelMKL MT2203 RNG ($64$ bit)\\	            
        \hline
                \Cref{fig:tpac_dt0-1} & IntelMKL MT2203 RNG ($64$ bit)\\	            
        \hline
                \Cref{fig:tpac_dt1zoom} & IntelMKL MT2203 RNG ($64$ bit)\\	            
              
        \hline
                         \Cref{fig:tpaci_dt1}& IntelMKL MT2203 RNG ($64$ bit)\\            
        \hline
                         \Cref{fig:tpaci_dt0-1} & IntelMKL MT2203 RNG ($64$ bit)\\            
        \hline
    \end{tabular}
    \caption{Random Generators for the Simulations.}
    \label{tab:RNGTable}
\end{table}

%-- LITERATUR ----------------------------------------------------------%
	%\clearpage
	%\addcontentsline{toc}{section}{Literaturverzeichnis}
	\bibliography{Quellen/SABCEMM.bib}
	\bibliographystyle{abbrvdin}

\end{document}